\documentclass{article}
\usepackage[11pt]{extsizes}
\usepackage[T2A]{fontenc}
\usepackage[utf8]{inputenc}
\usepackage{cmap}
\usepackage{slashed}
\usepackage{hyperref}
\usepackage{cite}
\usepackage{color}
\usepackage{graphicx}

\usepackage{amsmath}\numberwithin{equation}{section}
\usepackage{amsfonts}
\usepackage{mathrsfs}
\usepackage{amssymb}

\usepackage{ulem}

\usepackage{slashed}

\allowdisplaybreaks[1]

\begin{document}

\begin{titlepage}

\begin{center}
{\Large\bf On a quantization of deformed reducible gauge theories}

\vspace{1cm}
{\bf A.A. Averianov$^{1,3}$, A.O. Barvinsky$^{2,4}$, I.L. Buchbinder$^{3,5}$, V.A. Krykhtin$^{6}$, D.V. Nesterov$^{2}$}

\vspace{0.5cm}
{\it $^1$Moscow Institute of Physics and Technology,\\
141700 Dolgoprudny, Moscow region, Russia,}\\
{\tt averianov.aa@phystech.edu}\\
{\it $^2$I.E. Tamm Theory Department, P.N. Lebedev Physical Institute,\\
53 Leninsky Prospect, 119991, Moscow, Russia,\\
{\tt barvin@lpi.ru,\, nesterov@lpi.ru}\\
$^3$Bogoliubov Laboratory of Theoretical Physics,\\
 Joint Institute for Nuclear Research, 6, Joliot Curie, 141980 Dubna, Russia,\\
  {\tt buchbinder@theor.jinr.ru}\\
 $^4$Institute for Theoretical and Mathematical Physics,\\
Moscow State University, Leninskie Gory,
GSP-1, Moscow, 119991, Russia,\\
$^5$ Center of Theoretical Physics, Tomsk State Pedagogical University,\\
60 Kievskaya str., 634061, Tomsk, Russia\\
$^6$ National Research Tomsk Polytechnic University,\\
Lenin Av. 30, 634050, Tomsk, Russia,}\\
{\tt krykhin@tpu.ru}

\vspace{0.5cm}

\end{center}

\begin{abstract}
We consider a general reducible gauge theory deformed by mass or/and
interaction terms violating gauge invariance. It is shown that in the
Abelian case, by using the Stueckelberg-type procedure, this theory with
broken gauge symmetry can be converted into exactly gauge-invariant theory
which under a suitable choice of gauge conditions can be treated within
the formalism of minimal wave operators manageable by the
covariant Schwinger-DeWitt technique. We carry out quantization
of such a theory in general terms when the
initial generators of gauge transformations are of the first and second
stages of reducibility and derive its partition function in terms of the
functional integral with all corresponding ghost fields. This method
is applied to quantization of massive fermionic totally
antisymmetric tensor field models in $AdS$ space. One-loop
quantum effective action for these models is derived in the form of
the functional determinants of special Dirac-type differential operators in
various dimensions.
\end{abstract}

\end{titlepage}

\section{Introduction}
Gauge fields are the basic ingredients of all known approaches to unifying
fundamental interactions. Therefore, studying various aspects
related to quantization of gauge theories deserves special
attention. The emergence of new gauge field models naturally leads
to the problem of extending the quantization methods.

Discovery of supergravity in diverse dimensions and string/brane
models inspired interest in the study of field theories
containing antisymmetric tensor fields geometrically described by
$p$-forms (see, e.g., \cite{GSchW, P, J, O, BBSch, FVP, BLT, Ta}). A distinctive
feature of all such fields is that the corresponding gauge
transformations are significantly different from the ones in Yang-Mills
theory. It is known that massless $p$-form field models are the
reducible gauge theories related by a net of dualities. To be
more precise, in $d$ dimensions, the the $p$-form models and $d-p-2$
form models are classically equivalent (see, e.g., \cite{HL} and
references therein). This gives rise to a natural question about the
quantum equivalence of such models, which in turn requires their
quantization. Since, the gauge structure of $p$-form field models
are reducible gauge theories, the quantization cannot be carried on
the basis of generally accepted Faddeev-Popov and DeWitt
quantization rules applicable to Yang-Mills type fields.

In the context of duality between different gauge theories, there is
some interest in studying deformed $p$-form models, where a
gauge-invariant theory is deformed by mass terms or interaction
terms that violate gauge invariance\footnote{For example, in $d$
dimensions, the massive $p$-form field model is dual to $d-p-1$-form
field model (see, e.g., \cite{KT} and references therein).}. Such theories
arise within the context of cosmological applications
mainly due to the fact that they allow one to formulate new models with
nonminimal couplings to gravity (see, e.g., recent papers \cite{H},
\cite{HO} and references therein). In this context there
arises the problem of calculating quantum effective action in curved spacetime.
As is well known, the one-loop effective action is given by the functional
determinant of the wave operator in the quadratic part of the classical action. In
deformed gauge theories, the kinetic part of the action is gauge
invariant but the mass or interaction terms are not, so that the corresponding
differential operator becomes nonminimal. For such operators the manifestly
covariant Schwinger-DeWitt method of calculating the effective action
becomes not directly applicable\footnote{Problems of calculating effective action
associated with non-minimal operators are discussed e.g., in
\cite{BKW}.}. Therefore, it seems useful to develop a quantization scheme for
deformed theories that would allow one to circumvent the origin
of nonminimal operators and associated with them difficulties.

In principle, a simple and natural way to avoid a problem of
nonminimal operators in deformed gauge theories can be based on a
generalization of the Stueckelberg trick (see various
applications and uses of the Stueckelberg trick in massive 
Yang-Mills theories in the review \cite{RR}). The conventional
Stueckelberg trick allows one to convert non-gauge massive vector field
model into classically equivalent gauge model with initial gauge
symmetry. This is achieved by introducing an auxiliary scalar field
whose gauge transformation does not contain derivatives. Due to
gauge invariance we can now impose the same gauge fixing conditions
as in the massless vector field model and get a minimal differential
operator in the quadratic part of action. The price to be paid for that
is the necessity to use an additional auxiliary scalar field.
Generalization of such a trick for general deformed gauge theories would open  the
possibility to apply the covariant methods to calculate the
effective action in these theories. Such a procedure was
generalized for quantizing various bosonic (super) $p$-form
field models in curved spacetime and for studying the quantum duality
aspects (see, e.g., \cite{BBS, BKP, KT, BVOS}
and references therein\footnote{We especially want to emphasize the paper
\cite{KT} which presents a large number of new results and provides
a detailed list of publications concerning the quantization of
various massive bosonic and superfield $p$-form models.}). In this
paper we formulate the Stueckelberg trick for a
generic deformed Abelian gauge theory and then apply it for
quantization of the recently proposed massive totally antisymmetric
tensor spinor field model (fermionic $p$-form field model)
\cite{BKR}\footnote{In \cite{BKR} both massive and massless (gauge
invariant) fermionic $p$-models on $AdS_d$ space were constructed.
Massless fermionic $p$-models were also constructed in \cite{Z},
\cite{CFMS}. Some examples of fermionic $p$-form field models were
proposed in the context of six-dimensional exotic supergravity
\cite{H-1, H-2, H-3, W, B, HLL, HLMP, BHHS, Gu}.} along with the derivation of its effective actions.

Antisymmetric tensor field theories possess remarkable properties in
both classical and quantum domains, so the study of such
theories has attracted considerable attention. Here we focus only on
quantum aspects. The first model of antisymmetric tensor field in four
dimensions has been discovered by Ogievetsky and Polubarinov in 1967
\cite{OP} and then rediscovered in the context of string theory in
\cite{KR, CSc, GScO}\footnote{See also the discussion of
this model in \cite{Iv}.}. The four dimensional massive antisymmetric tensor field model has been proposed 
some earlier \cite{KUMM} where a duality of this model to massive vector field model was noted as 
well as. Later a large number of bosonic and
supersymmetric $p$-form field models in various dimensions were
proposed (see a complete enough list of corresponding references in
\cite{KT, KR}).

Quantization of the bosonic $p$-form field models were discussed by many
authors using various methods (see the pioneering papers
\cite{S1, S2, Sig, SN, HKO, Obukh}). Non-Abelian $p$-forms and their quantum structure
were considered in \cite{FT, TB, FrTs, AGM1, AF, BK1, AGM2, BG, N} (see also
the review \cite{GPS}). The construction and the analysis of quantum
structure of superfield models containing $p$-forms among their field
components were discussed e.g., in \cite{S1, G, GNSZ, BK2, BuKu, FINN, KT, KR}
\footnote{See a complete list of references on this issue
in \cite{KT, KR}.}.

All the $p$-form field theories considered above are reducible gauge
theories, which means that their generators of gauge
transformations are linearly dependent. Currently, there exist general
methods for quantizing almost every possible gauge theory---the canonical BFV method \cite{FV, BF} (see also the book \cite{Henn} and the
references therein) and the covariant BV method \cite{Bat1, Bat2}
(see also the review \cite{GPS}, the book \cite{Henn} and
references therein). However, application of general methods for
quantizing fairly simple and mostly Abelian p-form gauge theories is
unnecessarily cumbersome. General methods require the use of a large
number of auxiliary ghost fields, the exclusion of which in fairly
simple theories for the sake of obtaining the final result
may be requiring a lot of efforts. This was especially pointed out
in the pioneering BV paper \cite{Bat2}. However, quite a
long time ago, in the work of \cite{BK2} (see also \cite{BuKu}) in the context of studying quantum
effective actions for dual superfield theories on a supergravity
background, a simple quantization procedure was proposed. Recently
this procedure was generalized for quantization of general reducible
gauge theories \cite{BBKN1}. Such a procedure is ideally acceptable
for quantization of $p$-form gauge theories that was demonstrated in
\cite{BBKN1} on the example of quantizing the fermionic $p$-form theory on
$AdS$ background. It was also shown that the BV quantization method
leads to the same results \cite{BBKN2}. BV quantization of the
fermionic $p$-form theory on a flat background was carried out in
\cite{LZ}.

Quantization of deformed $p$-form field theories is not so well
developed as for the  massless $p$-form field theories. As far as we know,
there are only few papers on this subject (see, e.g., \cite{BKP, KT},
and references therein). In this paper, we formulate the generic Stuckelberg
trick for arbitrary deformed Abelian reducible gauge theories
converting them into equivalent
purely gauge theories which include Stueckelberg fields of a generic type. Quantization of
these theories is carried out on the basis of the procedure proposed in
\cite{BBKN1}, and one can expect that gauge
transformations of Stueckelberg fields may turn out to be reducible
as well. As an application of this approach we derive the effective
action for massive fermionic $p$-form theories of first and second
stages of reducibility on the $AdS$ background.

The paper is organized as follows. In section \ref{Sect:deformed_theories} we describe the
deformation of a general bosonic gauge theory and show how this theory
can be converted into purely gauge theory on the basis of the
Stueckelberg trick formulated for an Abelian general gauge
theory. Section \ref{Sect:first-stage} is devoted to the quantization of a deformed gauge theory
in the case of the first stage of reducibility by using the procedure
of \cite{BBKN1}. In section \ref{Sect:second-stage} we apply the results of
section \ref{Sect:first-stage} to the deformed theories of the second stage of reducibility
again by using the quantization procedure of \cite{BBKN1}. Here we show
that in this case the gauge transformations of the Stueckelberg
fields are themselves of the first stage of reducibility. Section \ref{Sect:spin-tensor} is devoted to the quantization of massive antisymmetric tensor
spinor theories for the first and second stages of reducibility\footnote{To quantize these models, we use the results of the previous sections obtained for bosonic fields. Since the fermionic p-form models are described by anticommuting fields, we apply the quantization procedure described above, but unlike the previous sections, we employ here the functional integrals over anicommuting variables.}. The results are the effective actions written in terms
of functional determinants of special differential operators acting
on fermionic $p$-forms. In section \ref{Sect:conclusions} we give our concluding remarks.

\newcommand{\HIDE}[1]{}
\newcommand{\DN}[1]{{\color{blue}#1}}

\section{Deformed gauge theories}
 \label{Sect:deformed_theories}
\subsection{Deformation of general gauge theory}
General gauge theory
is given by the set of fields $\phi^{i}$, their action $S_0[\phi]$,
infinitesimal gauge transformations $\delta\phi^{i} =
R^{i}_{\alpha}[\phi]f^{\alpha}$ and the condition of gauge invariance
\begin{equation}
\label{invariance}
S_{0}[\phi]_{,i}R^{i}_{\alpha}[\phi]=0.
\end{equation}
The functionals $R^{i}_{\alpha}[\phi]$ are called the generators of gauge transformations or gauge generators and $f^{\alpha}$ are the gauge parameters\footnote{We use DeWitt condensed notations where the indices $i$ and $\alpha$ contain all continuous and discrete labels of fields and parameters. For contracted indices, the integration (their continuous part) and summation (over their discrete part) is assumed. $A_{,i}[\phi] = \delta{A[\phi]}/\delta{\phi^{i}}$ for any functional $A[\phi]$.}. The fields $\phi^{i}$ can possess non-zero Grassmann parity, however, for simplicity, here we will consider only bosonic fields and call the corresponding theory a general bosonic gauge theory.

The gauge theory is called reducible if the gauge generators are linearly dependent, which means that there exist the functionals $Z^{(1)}{}^{\alpha}_{\,a_1}[\phi]$ enumerated by the indices $a_1$ such that
\begin{equation}
\label{dependence}
R^{i}_{\alpha}[\phi]Z^{(1)}{}^{\alpha}_{\,a_1}=0 \qquad \alpha=1,...\,m_0,\qquad a_1= {1,...\, m_1 < m_0}.
\end{equation}
In the opposite case it is called irreducible. The relation (\ref{dependence}) means that in reducible theories, the $Z^{(1)}{}^{\alpha}_{\,a_1}[\phi]$ are the zero vectors of gauge generators. Here $m_0$ and $m_1$ denote respectively formal dimensionalities of the original gauge algebra and the space of $Z^{(1)}{}^{\alpha}_{\,a_1}$, the latter being linearly independent for stage one reducible gauge theories. This means that the actual number of local gauge symmetries is $m_0-m_1$ (the rank of the matrix $R^i_\alpha[\phi]$) rather than $m_0$. Therefore the actual number of independent gauge conditions to fix these symmetries should also be $m_0-m_1$. They can be represented by the redundant (or reducible) set of gauge conditions $\chi^{\mu_1}$ satisfying the condition of linear dependence with some zero vectors $\bar Z^{(1)}{}^{b_1}_{\mu_1}$,
\begin{equation}\label{left_zero}
\bar Z^{(1)}{}^{b_1}_{\mu_1}\chi^{\mu_1}= 0,\qquad \mu_1=1,...\,m_0,\qquad b_1= {1,...\, m_1 < m_0}.
\end{equation}

If the matrix $Z^{(1)}{}^{\alpha}_{\,a_1}$ has no zero vectors, the corresponding theory is called the one of the first-stage of reducibility. If it has zero vectors
$Z^{(2)}{}^{a_1}_{\,a_2}$ so that
\begin{equation}
Z^{(1)}{}^{\alpha}_{\,a_1} Z^{(2)}{}^{a_1}_{\,a_2}=0,  \qquad a_2 = {1,...\, m_2 < m_1}
\end{equation}
and $Z^{(2)}{}^{a_1}_{\,a_2}$ has no zero vectors, the corresponding theory is called the gauge theory of the second stage of reducibility. The theories of higher stages of reducibility are defined analogously.

Consider a theory with action
\begin{equation}
\label{deformed}
S[\phi] = S_0[\phi] + S_1[\phi],
\end{equation}
where $R^{i}_{\alpha}S_{,i}= R^{i}_{\alpha}S_{1\, ,i} \neq 0$. This relation shows that the theory under consideration is non gauge. Since the theory with the action (\ref{deformed}) is obtained by deforming the initial gauge theory, we will call such a theory the deformed gauge theory. As to the action $S_1[\phi]$, we will assume that it has the following form
\begin{equation}
\label{deformstruct}
S_1[\phi] = \frac{1}{2}t_{ij}\phi^{i}\phi^{j} + \frac{1}{3!}t_{ijk}\phi^{i}\phi^{j}\phi^{k} + \frac{1}{4!}t_{ijkl}\phi^{i}\phi^{j}\phi^{k}\phi^{l}+ \ldots
\end{equation}
with $t_{ij},\,t_{ijk},\,t_{ijkl},\,\ldots$ being its deformation parameters. Since the initial theory can be either irreducible or reducible, we will use for the corresponding deformed theory the name of irreducible or reducible gauge theory.

Examples of deformed gauge theories are as follows.
\begin{itemize}
\item{Non-reducible deformed vector field model with the action:}
$$
S[A]= \int d^4 x\sqrt{|g|} \big(-\frac{1}{4} F_{\mu\nu}F^{\mu\nu} - \frac{m^2}{2} A^{\mu}A_{\mu} -\frac{\lambda}{4!} (A^{\mu}A_{\mu})^2\big).
$$
Here $F_{\mu\nu}$ is the standard field strength tensor for a vector field $A_{\mu}$ and $m^2$ and $\lambda$ are the deformation parameters.
\item{Deformed second rank antisymmetric tensor field theory with the action (the first-stage deformed gauge theory):}
$$
S[B]= \int d^4x \sqrt{|g|}\big(-\frac{1}{12}F_{\mu\nu\rho}F^{\mu\nu\rho}-\frac{m^2}{2}B_{\mu\nu}B^{\mu\nu} -\frac{\lambda}{4!}(B_{\mu\nu}B^{\mu\nu})^2\big).
$$
Here $F_{\mu\nu\rho}= \nabla_{\mu}B_{\nu\rho}+\nabla_{\rho}B_{\mu\nu}+\nabla_{\nu}B_{\rho\mu}$ is the field strength tensor for the second rank antisymmetric tensor field $B_{\mu\nu},$  and the $m^2$ and $\lambda$ are the deformation parameters.
\item{Deformed totally antisymmetric tensor spinor field model (higher stage deformed gauge theory). This teory is defined only on $AdS_d$ space and is described by the action:}
$$
S[\psi,\bar{\psi}] = \int d^px \sqrt{|g|}\big(\bar{\psi}_{\mu_1\mu_2 \ldots \mu_p}i\gamma^{\mu_1\mu_2 \ldots \mu_p\sigma\nu_1\nu_2 \ldots \nu_p}D_{\sigma}
\psi_{\nu_1\nu_2 \ldots \nu_p} - m \bar{\psi}_{\mu_1\mu_2 \ldots \mu_p}\gamma^{\mu_1\mu_2 \ldots \mu_p\nu_1\nu_2 \ldots \nu_p}\psi_{\nu_1\nu_2 \ldots \nu_p}\big)
$$
Here $\psi_{\mu_1\mu_2 \ldots \mu_p}=
\psi_{[\mu_1\mu_2 \ldots \mu_p]}=\psi_{\mu[p]}$ is a fermionic $p$-form and $\mu[p]=[\mu_1\mu_2 \ldots \mu_p]$ implies the collection of $p$ antisymmetric indices, $D_\mu = \nabla_{\mu} \pm \frac{i}{2}\sqrt{r}\gamma_{\mu}$, $\nabla_{\mu}$ is the spinor covariant derivative, $r$ is  determined from $R^{\mu\nu}{}_{\rho\sigma}=r(\delta^\mu_\sigma\delta^\nu_\rho-\delta^\mu_\rho\delta^\nu_\sigma)$ and $r>0$ for the $AdS$ space, and $m$ is a deformation mass parameter.
At $m=0$, the action is invariant under the transformations
$$
\delta{\psi}_{\mu[p]}= pD_{\mu}\lambda_{\mu[p-1]},
\quad
\delta{\lambda}_{\mu[p-1]}=(p-1)D_{\mu}\lambda_{\mu[p-2]},
\quad
\ldots
$$
where ${\lambda}_{\mu[k]},\, k=0,1, \ldots p-1$ are totally anticommuting rank $k$ tensor spinor gauge parameters. In the massless case, the theory under consideration possesses $p-1$ stages of reducibility.
\end{itemize}

\subsection{Conversion of a deformed Abelian gauge theory into a
complete gauge invariant theory}

Consider the Abelian gauge theory with the action $S_0[\phi]$ and field independent gauge generators $R^{i}_{\alpha}$. Such gauge generators are typical for
antisymmetric field models. The action $S_0[\phi]$ is deformed by the term $S_1[\phi]$, where $S_1[\phi]$ describes deformation of the form (\ref{deformstruct}). As a result we arrive at the theory with action $S[\phi]=S_0[\phi]+S_1[\phi],$ where the first term is the gauge invariant initial action and the second term breaks gauge invariance.

Let us introduce the field $\hat{\phi}^i$ by using the generic Stuekelberg trick
\begin{equation}
\label{hat}
\hat{\phi}^i = \phi^i - R^{i}_{\alpha}s^{\alpha} = \hat{\phi}^{i}(\phi,s),
\end{equation}
where $s^{\alpha}$ are some new fields. It is evident that the field $\hat{\phi}^i$ is invariant under the following gauge transformations
\begin{equation}
\label{gauge-st}
\delta{\phi}^i= R^{i}_{\alpha}f^{\alpha}, \qquad \delta{s}^{\alpha} = f^{\alpha}.
\end{equation}
Consider a theory with the action $\hat{S[\phi,s]}$ in the form
\begin{equation}
\label{action-st}
\hat{S}[\phi,s] = S_0[\phi] + S_1[\hat{\phi}(\phi,s)].
\end{equation}
It is easy to see that the action (\ref{action-st}) is gauge invariant under the transformations $\delta{\phi}^i = R^{i}_{\alpha}f^{\alpha},\, \delta s^{\alpha}=f^{\alpha}$. Due to this gauge invariance, we can impose the gauge conditions number of which coincides with a number of gauge parameters $f^{\alpha}$. The simplest such gauge conditions are $s^{\alpha}=0.$ The parameters $f^{\alpha}$ are completely fixed and the gauge invariant action (\ref{action-st}) reduces to deformed action (\ref{deformed}). These consideration show that the theories with actions (\ref{action-st}) and (\ref{deformed}) are equivalent as the classical field models. Taking into account all above, it is natural to call the fields $s^{\alpha}$, the general Stueckelberg fields\footnote{
 We note that the Stueckelberg construction presented here is strictly valid only for Abelian gauge symmetries, where the generators of gauge transformations are field-independent. For a general deformed gauge theory, introducing Stueckelberg fields requires a significantly more involved procedure. However, since this work focuses exclusively on deformed reducible gauge models based on bosonic and fermionic $p$-forms---whose gauge transformations remain Abelian---the proposed approach is adequate and sufficient.
}.

Now, taking into account the invariance of the action (\ref{action-st}), we can use gauge conditions which will fix the gauge structure of the initial gauge invariant part $S_0[\phi]$ of the full deformed action. In the case under consideration, the kinetic terms in the deformed action (\ref{deformed}) are related only to this gauge invariant part $S_0[\phi]$. Therefore, in the context of calculating effective action it is more appropriate to impose the gauge fixing condition not on the fields $s^{\alpha}$ but on the initial fields ${\phi}^i$ to simplify their kinetic terms. As a result, this can allow one to avoid nonminimal operators, which is exactly what we will do when quantizing the massive totally antisymmetric tensor spinor theory in section \ref{Sect:spin-tensor}. 

\section{Quantization of first-stage reducible deformed gauge theories}
 \label{Sect:first-stage}
\subsection{Modified delta functions and group integration measure}
Now consider the first-stage reducible Abelian theory with the action \eqref{action-st}. As noted above, in this case the matrix of gauge generators $R^{i}_{\alpha}$ admits a set of linearly independent zero vectors ${Z^{(1)}}{}^{\alpha}_{a_{1}}$. It follows that the field $\hat{\phi}^{i}$ introduced in \eqref{hat} is invariant under the transformation $\delta s^{\alpha}={Z^{(1)}}{}^{\alpha}_{a_{1}}g^{a_{1}}$, where $g^{a_{1}},\,a_{1}=1,...\,m_{1}$ are additional gauge parameters. As a result, the general gauge symmetry of the model is characterized by the set of $m_{0}$ independent parameters $f^{\alpha}$, the set of $m_1$ parameters $g^{a_{1}}$ and takes the form
\begin{equation}
\label{gauge-st-1st}
\delta{\phi}^i= R^{i}_{\alpha}f^{\alpha}, \qquad \delta{s}^{\alpha} = f^{\alpha}+{Z^{(1)}}{}^{\alpha}_{a_{1}}g^{a_{1}}.
\end{equation}
It is also easy to see that this transformation law is itself invariant under the transformation
\begin{equation}
\label{eq:5}
\delta f^{\alpha}={Z^{(1)}}{}^{\alpha}_{a_{1}}{f_{1}}^{a_{1}},\qquad\delta g^{a_{1}}=-{f_{1}}^{a_{1}},
\end{equation}
where ${f_{1}}^{a_{1}}$ is a set of $m_{1}$ second-stage gauge parameters.

The standard Faddeev-Popov procedure becomes inapplicable for quantization of theories with linearly dependent gauge generators. In \cite{BBKN1} the proper functional integral for a massless theory with first-stage reducibility is constructed by introducing according to the Faddeev-Popov trick \cite{FP} the additional unity into the expressions for the delta function of gauge conditions and group integration measure, and by separating integration over the second-stage gauge group. We apply this method to the quantization of massive theory with first-stage reducibility by using the Stueckelberg fields. The main difficulty in applying the Faddeev-Popov method to reducible gauge theories is that the factors in the integrand of the functional representation of the Faddeev-Popov unity
\begin{equation}
\label{naive-unit-1}
1=\Delta\int \mathscr{D}f\mathscr{D}g\;\delta[\chi^{\alpha}(\phi^{i}+R^{i}_{\beta}f^{\beta})]\;\delta[\omega^{a_{1}}_{\alpha}(s^{\alpha}+f^{\alpha}+{Z^{(1)}}{}^{\alpha}_{b_{1}}g^{b_{1}})],
\end{equation}
are degenerate (here $\chi^{\alpha}(\phi^{i})$ and $\omega^{a_{1}}_{\alpha}s^{\alpha}$ are two gauge-fixing conditions corresponding to the two symmetries \eqref{gauge-st}, with the rank of the gauge-fixing matrix $\omega^{a_{1}}_{\alpha}$ equal to $m_{1}$). Since the gauge symmetry of the theory can be parametrized by $m_{0}-m_{1}$ independent parameters $f$ and $m_{1}$ independent parameters $g$, one must impose $m_{0}$ independent gauge conditions. Because the conditions $\omega^{a_{1}}_{\alpha}s^{\alpha}$ are independent by construction, the conditions $\chi^{\alpha}(\phi^{i})$ must be linearly dependent, which implies the existence of the matrix of zero vectors ${\bar{Z}^{(1)}}{}^{a_{1}}_{\alpha}$ such that ${\bar{Z}^{(1)}}{}^{a_{1}}_{\alpha}\chi^{\alpha}(\phi^{i})=0$. As a result, the naive delta function
\begin{equation}
\label{eq:8}
\delta[\chi^{\alpha}(\phi^{i})^{f}]=\int \mathscr{D}\pi\;e^{i\pi_{\alpha}\chi^{\alpha}[(\phi^{i})^{f}]},
\end{equation}
where the superscript $f$ denotes the gauge transformation with the gauge parameter $f$, is degenerate due to invariance of the exponential under the transformation $\delta\pi_{\alpha}=\xi_{a_{1}}{\bar{Z}^{(1)}}{}_{\alpha}^{a_{1}}$. In this sense, the matrix ${\bar{Z}^{(1)}}{}_{\alpha}^{a_{1}}$ becomes the matrix of generators of second-stage gauge transformations. To lift this degeneracy, we fix the symmetry $\delta\pi_{\alpha}=\xi_{a_{1}}{\bar{Z}^{(1)}}{}_{\alpha}^{a_{1}}$ by imposing on $\pi^{\alpha}$ the gauge condition $\sigma(\pi)=\pi_{\alpha}\sigma_{a_{1}}^{\alpha}=0$ and insert into the expression \eqref{eq:8},  according to the Faddeev-Popov method, the functional unity
\begin{equation}
 1=\text{Det}({\bar{Z}^{(1)}}{}^{a_{1}}_{\alpha}\sigma^{\alpha}_{b_{1}})\int \mathscr{D}\xi\;
 \delta[(\pi_{\alpha}+\xi_{b_{1}}{\bar{Z}^{(1)}}{}^{b_{1}}_{\alpha})\sigma^{\alpha}_{a_{1}}]
 .
\end{equation}
Next, by changing the order of integration and using the invariance of measure $\mathscr{D}\pi$ under the transformation $\delta\pi_{\alpha}={\bar{Z}^{(1)}}{}_{\alpha}^{a_{1}}\xi_{a_{1}}$, one disentangles a (formally infinite) volume of the gauge group $\int \mathscr{D}\xi=V_{\xi}$.
\begin{align}
    \begin{gathered}
        \label{eq:15}
        \int \mathscr{D}\pi\;e^{i\pi_{\alpha}\chi^{\alpha}((\phi^{i})^{f})}\rightarrow\text{Det}({\bar{Z}^{(1)}}{}^{a_{1}}_{\alpha}\sigma^{\alpha}_{b_{1}})\int \mathscr{D}\pi\;e^{i\pi_{\alpha}\chi^{\alpha}((\phi^{i})^{f})}\int \mathscr{D}\xi\;
        \delta[(\pi_{\alpha}+\xi_{b_{1}}{\bar{Z}^{(1)}}{}^{b_{1}}_{\alpha})\sigma^{\alpha}_{a_{1}}]
=
\\
=
\text{Det}({\bar{Z}^{(1)}}{}^{a_{1}}_{\alpha}\sigma^{\alpha}_{b_{1}})\int \mathscr{D}\xi
\int\mathscr{D}\pi\; e^{i\pi_{\alpha}\chi^{\alpha}((\phi^{i})^{f})}
 \delta[\pi_{\alpha}\sigma^{\alpha}_{a_{1}}]
=
\\
=
V_{\xi}\times\text{Det}({\bar{Z}^{(1)}}{}^{a_{1}}_{\alpha}\sigma^{\alpha}_{b_{1}})
\int\mathscr{D}\pi\mathscr{D}c\; e^{i\pi_{\alpha}\chi^{\alpha}((\phi^{i})^{f})
 +i\pi_{\alpha}\sigma^{\alpha}_{a_{1}}c^{a_{1}}} 
=
\\
=
V_{\xi}\times \text{Det}
 ({\bar{Z}^{(1)}}{}^{a_{1}}_{\alpha}\sigma^{\alpha}_{b_{1}})
\int\mathscr{D}c\;\delta[\chi^{\alpha}((\phi^{i})^{f})+\sigma^{\alpha}_{a_{1}}c^{a_{1}}].
    \end{gathered}
\end{align}
As a result, one obtains the expression for a modified delta function, which should be used in construction of the Faddeev-Popov functional unity:
\begin{equation}
 \label{eq:16}
 \delta[\chi^{\alpha}(\phi^{i}+R^{i}_{\alpha}f^{\alpha})] \rightarrow \hat{\delta}[\chi^{\alpha}(\phi^{i}+R^{i}_{\alpha}f^{\alpha})]
 = \text{Det}
 ({\bar{Z}^{(1)}}{}^{a_{1}}_{\alpha}\sigma^{\alpha}_{b_{1}})
 \int\mathscr{D}c\;\delta[\chi^{\alpha}(\phi^{i}+R^{i}_{\alpha}f^{\alpha})+\sigma^{\alpha}_{a_{1}}c^{a_{1}}].
\end{equation}
The delta function for the Stueckelberg field is well defined (since the rank of $\omega^{a_{1}}_{\alpha}$ equals $m_{1}$) and does not require any modification.

Next, the integration measure over gauge parameters must be modified, $\mathscr{D}f\mathscr{D}g\rightarrow\mathscr{D}\mu_{f,g}$. The invariance of the integrand in the functional unity \eqref{naive-unit-1} under the transformations \eqref{eq:5} (now playing the role of second-stage gauge transformations) leads to the degeneracy of the integral. We fix the symmetry \eqref{eq:5} by imposing on integration variable $f$ the gauge condition $\omega(f)=\omega^{a_{1}}_{\alpha}f^{\alpha}=0$ (the same condition used for fixing Stueckelberg gauge transformations) and insert the functional unity into integral \eqref{naive-unit-1},
\begin{equation}
1=\text{Det}(\omega^{a_{1}}_{\alpha}{Z^{(1)}}{}^{\alpha}_{b_{1}})\int\mathscr{D}f_{1}\;\delta[\omega^{a_{1}}_{\alpha}(f^{\alpha})^{f_{1}}],
\end{equation}
where again the superscript $f_1$ denotes the gauge transformation \eqref{eq:5} with the parameter $f_1$. Then we perform a chain of transformations analogous to \eqref{eq:15} (denoting the integrand in \eqref{naive-unit-1} as $\Phi[\phi^{f},s^{f,g}]$ with $s^{f,g}$ obviously denoting the transformation (\ref{gauge-st-1st}) of $s$ under the action of $f$ and $g$ parameters):
\begin{align}
    \begin{gathered}
        \label{eq:18}
        \int\mathscr{D}f\mathscr{D}g\;\Phi[\phi^{f},s^{f,g}]\rightarrow\text{Det}(\omega^{a_{1}}_{\alpha}{Z^{(1)}}{}^{\alpha}_{b_{1}})\int\mathscr{D}f\mathscr{D}g\;\Phi[\phi^{f},s^{f,g}]\int\mathscr{D}f_{1}\;\delta[\omega^{a_{1}}_{\alpha}(f^{\alpha})^{f_{1}}]=\\
        = \text{Det}
        (\omega^{a_{1}}_{\alpha}{Z^{(1)}}{}^{\alpha}_{b_{1}})
        \int\mathscr{D}f_{1}\int\mathscr{D}f\mathscr{D}g\;\delta[\omega^{a_{1}}_{\alpha}f^{\alpha}]\;\Phi[\phi^{f},s^{f,g}]
        =\\
        =V_{f_{1}}\times
        \text{Det}
         (\omega^{a_{1}}_{\alpha}{Z^{(1)}}{}^{\alpha}_{b_{1}})
        \int\mathscr{D}f\mathscr{D}g\;\delta[\omega^{a_{1}}_{\alpha}f^{\alpha}]\;\Phi[\phi^{f},s^{f,g}]
    \end{gathered}
\end{align}
As a result, one obtains the expression for the modified integration measure,
\begin{equation}
\label{eq:19}
  \mathscr{D}f\mathscr{D}g
  \rightarrow
  \mathscr{D}\mu_{f,g}
  =\mathscr{D}f\mathscr{D}g\;
  \text{Det}
     (\omega^{a_{1}}_{\alpha}{Z^{(1)}}{}^{\alpha}_{b_{1}}) \; \delta[\omega^{a_{1}}_{\alpha}f^{\alpha}].
\end{equation}

\subsection{Faddeev-Popov determinant and functional integral}
In order to compute the Faddeev–Popov determinant, we substitute expressions \eqref{eq:16}, \eqref{eq:19} in the integral \eqref{naive-unit-1} and transfer all the determinant factors to the left hand side of \eqref{naive-unit-1}, raised to the power $-1$:
\begin{align}
    \begin{gathered}
        \Delta^{-1}\;\text{Det}^{-1}(\omega^{a_{1}}_{\alpha}{Z^{(1)}}{}^{\alpha}_{b_{1}})\;
        \text{Det}^{-1}({\bar{Z}^{(1)}}{}^{a_{1}}_{\alpha}\sigma^{\alpha}_{b_{1}})=\\
        =\int\mathscr{D}f\mathscr{D}g\mathscr{D}c\;\delta[\omega^{a_{1}}_{\alpha}f^{\alpha}]\;\delta[\chi^{\alpha}(\phi^{i}+R^{i}_{\beta}f^{\beta})+\sigma^{\alpha}_{a_{1}}c^{a_{1}}]\;\delta[\omega^{a_{1}}_{\alpha}(s^{\alpha}+f^{\alpha}+{Z^{(1)}}{}^{\alpha}_{b_{1}}g^{b_{1}})]
=
\\
=
\text{Det}^{-1}\begin{pmatrix}\frac{\delta\chi^{\alpha}}{\delta\phi^{i}}R^{i}_{\beta} & \sigma^{\alpha}_{b_{1}} & 0 \\ \omega^{a_{1}}_{\beta} & 0 & 0 \\ \omega^{c_{1}}_{\beta} & 0 & \omega^{c_{1}}_{\gamma}{Z^{(1)}}{}^{\gamma}_{d_{1}}\end{pmatrix}=\text{Det}^{-1}\begin{pmatrix}\frac{\delta\chi^{\alpha}}{\delta\phi^{i}}R^{i}_{\beta} & \sigma^{\alpha}_{b_{1}}\\ \omega^{a_{1}}_{\beta} & 0\end{pmatrix}\text{Det}^{-1}(\omega^{c_{1}}_{\gamma}{Z^{(1)}}{}^{\gamma}_{d_{1}})\label{eq:20}
    \end{gathered}
\end{align}
In \cite{BBKN1} it was shown that
\begin{equation}
\label{eq:21}
\text{Det}\begin{pmatrix}\frac{\delta\chi^{\alpha}}{\delta\phi^{i}}R^{i}_{\beta} & \sigma^{\alpha}_{b_{1}}\\ \omega^{a_{1}}_{\beta} & 0\end{pmatrix}=\text{Det}(F^{\alpha}_{\beta}),
\end{equation}
where $F^{\alpha}_{\beta}=\frac{\delta\chi^{\alpha}}{\delta\phi^{i}}R^{i}_{\beta}+\sigma^{\alpha}_{a_{1}}\omega^{a_{1}}_{\beta}$. Using this relation one obtains the yet unknown expression for the Faddeev–Popov determinant $\Delta$,
\begin{equation}
\Delta=\frac{\text{Det}(F^{\alpha}_{\beta})}
 {
   \text{Det}({\bar{Z}^{(1)}}{}^{a_{1}}_{\alpha}\sigma^{\alpha}_{b_{1}})
 }.
\end{equation}
Now we insert into the functional integral the correct unity expression
\begin{equation}
1=\Delta\int \mathscr{D}\mu_{f,g}\;\hat{\delta}[\chi^{\alpha}(\phi^{i}+R^{i}_{\beta}f^{\beta})]\;
\delta[\omega^{a_{1}}_{\alpha}(s^{\alpha}+f^{\alpha}+{Z^{(1)}}{}^{\alpha}_{b_{1}}g^{b_{1}})],
\end{equation}
where the first (hatted) delta function is given by the functional integral representation (\ref{eq:16}),
and perform the same sequence of transformations as in \eqref{eq:15}, \eqref{eq:18}:
\begin{align}
    \begin{gathered}
        \label{eq:24}
        \int\mathscr{D}\phi\mathscr{D}s\;e^{iS[\phi,s]}\rightarrow\\
        \rightarrow\text{Det}(F)\int\mathscr{D}\phi\mathscr{D}s\;e^{iS[\phi,s]}\int \mathscr{D}\mu_{f,g}\int\mathscr{D}c\;\delta[\chi^{\alpha}(\phi^{i}+R^{i}_{\alpha}f^{\alpha})+\sigma^{\alpha}_{a_{1}}c^{a_{1}}]\;\delta[\omega^{a_{1}}_{\alpha}(s^{\alpha}+f^{\alpha}+{Z^{(1)}}{}^{\alpha}_{b_{1}}g^{b_{1}})]=\\
        =V\times\text{Det}(F)\int\mathscr{D}\phi\mathscr{D}s\mathscr{D}c\;e^{iS[\phi,s]}\;\delta[\chi^{\alpha}(\phi^{i})+\sigma^{\alpha}_{a_{1}}c^{a_{1}}]\;\delta[\omega^{a_{1}}_{\alpha}s^{\alpha}]
    \end{gathered}
\end{align}
As a result, one arrives at the following final expression for the functional integral of a massive first-stage reducible gauge theory with Stueckelberg fields
\begin{equation}
Z=\text{Det}(F)\int\mathscr{D}\phi\mathscr{D}s\mathscr{D}c\;e^{iS[\phi,s]}\;\delta[\chi^{\alpha}(\phi^{i})+\sigma^{\alpha}_{a_{1}}c^{a_{1}}]\;\delta[\omega^{a_{1}}_{\alpha}s^{\alpha}]
\end{equation}
Due to the choice of gauge-fixing conditions diagonal with respect to the fields $\phi$ and $s$, the delta functions also turns out to be diagonal (containing either only $\phi$ or only $s$ fields). However, this does not imply that fields $\phi$ and $s$ are completely decoupled, since their mixing remains in $S[\phi,s]$.

\section{Quantization of deformed second-stage reducible gauge theory}
 \label{Sect:second-stage}
\subsection{Modified group integration measure}

In the case of second-stage reducibility expressions the transformations \eqref{gauge-st-1st} are themselves invariant under the additional gauge transformation $\delta g^{a_{1}}={Z^{(2)}}{}^{a_{1}}_{a_{2}}{g_{1}}^{a_{2}}$. Therefore, the general second-stage gauge transformation takes the form
\begin{equation}
\label{eq:4}
\delta f^{\alpha}= {Z^{(1)}}{}^{\alpha}_{a_{1}}{f_{1}}^{a_{1}}\qquad
\delta g^{a_{1}}= -{f_{1}}^{a_{1}}+{Z^{(2)}}{}^{a_{1}}_{a_{2}}{g_{1}}^{a_{2}},
\end{equation}
where ${g_{1}}^{a_{2}}$ are additional second-stage gauge parameters. By analogy with the first-stage reducibility, the transformation law \eqref{eq:4} is invariant under the transformation
\begin{equation}
\label{eq:6}
\delta{f_{1}}^{a_{1}}={Z^{(2)}}{}^{a_{1}}_{a_{2}}{f_{2}}^{a_{2}}\qquad
\delta{g_{1}}^{a_{2}}={f_{2}}^{a_{2}},
\end{equation}
where ${f_{2}}^{a_{2}}$ are third-stage gauge parameters.

In order to consistently construct the functional integral of a second-stage reducible gauge theory, it is necessary, similarly to the previous case, to define modified delta functions and integration measure and to insert into the functional integral the following unity expression:
\begin{equation}
\label{eq:26}
1=\Delta\int \mathscr{D}\mu_{{f,g}}\;\hat{\hat{\delta}}[\chi^{\alpha}(\phi^{i}+R^{i}_{\alpha}f^{\alpha})]\;\hat{\delta}[\omega^{a_{1}}_{\alpha}(s^{\alpha}+f^{\alpha}+{Z^{(1)}}{}^{\alpha}_{b_{1}}g^{b_{1}})]
\end{equation}
Here $\chi^{\alpha}$ denotes gauge fixing for fields $\phi$, while the matrix $\omega^{a_{1}}_{\alpha}$ fixes the gauge for Stueckelberg fields $s$.\

First, we determine the modified integration measure $\mathscr{D}\mu_{f,g}$. A naive integral
\begin{equation}
\label{eq:27}
1=\Delta\int \mathscr{D}f\mathscr{D}g\;\delta[\chi^{\alpha}(\phi^{i}+R^{i}_{\alpha}f^{\alpha})]\;\delta[\omega^{a_{1}}_{\alpha}(s^{\alpha}+f^{\alpha}+{Z^{(1)}}{}^{\alpha}_{b_{1}}g^{b_{1}})],
\end{equation}
turns out to be degenerate due to the invariance of the expressions inside delta functions with respect to transformation \eqref{eq:4}. To fix this symmetry, we choose gauge conditions
\begin{equation}
\omega(f)=\omega^{a_{1}}_{\alpha}f^{\alpha}\;\;\;\;\;\omega'(g)={\omega'}^{a_{2}}_{a_{1}}g^{a_{1}},
\end{equation}
and insert into the integral \eqref{eq:27} the functional unity
\begin{equation}
\label{eq:29}
1=\Delta_{m}\int \mathscr{D}\mu_{{f_{1},g_{1}}}\;\hat{\delta}[\omega^{a_{1}}_{\alpha}(f^{\alpha}+{Z^{(1)}}{}^{\alpha}_{b_{1}}{f_{1}}^{b_{1}})]\;\delta[{\omega'}^{a_{2}}_{a_{1}}(g^{a_{1}}-{f_{1}}^{a_{1}}+{Z^{(2)}}{}^{a_{1}}_{b_{2}}{g_{1}}^{b_{2}})]
\end{equation}
In this expression the symmetry \eqref{eq:6} plays the role of third-stage gauge transformations, so that in complete analogy with the previous case one has to define the modified measure. In addition, since the matrix $\omega$ now possesses $m_{2}$ independent null vectors, the exponential in the expression
\begin{equation}
\delta[\omega^{a_{1}}_{\alpha}(f^{\alpha}+{Z^{(1)}}{}^{\alpha}_{b_{1}}{f_{1}}^{b_{1}})]=\int \mathscr{D}\pi\;e^{i\pi_{a_{1}}\omega^{a_{1}}_{\alpha}(f^{\alpha}+{Z^{(1)}}{}^{\alpha}_{b_{1}}{f_{1}}^{b_{1}})}
\end{equation}
is invariant under the transformation $\delta\pi_{a_{1}}=\xi_{a_{2}}{{\tilde{Z}^{(2)}}}{}_{a_{1}}^{a_{2}}$, where ${{\tilde{Z}^{(2)}}}$ is the matrix (of rank $m_{2}$) of null vectors of the gauge condition $\omega(f)$. This leads to the necessity of defining the modified delta function. This symmetry is fixed by the gauge condition $\sigma'{}(\pi)=\pi_{a_{1}}\sigma'{}^{a_{1}}_{a_{2}}$, while symmetry \eqref{eq:6} is fixed by the condition $\omega'(f_{1})=\omega'{}_{a_{1}}^{a_{2}}{f_{1}}^{a_{1}}$ (rank of the matrices $\sigma'{}$ and $\omega'$ is equal to $m_{2}$).\
Similarly to the previous case, we insert into a naive integral
\begin{equation}
1=\Delta_{m}\int \mathscr{D}f_{1}\mathscr{D}g_{1}\;\delta[\omega^{a_{1}}_{\alpha}(f^{\alpha}+{Z^{(1)}}{}^{\alpha}_{b_{1}}{f_{1}}^{b_{1}})]\;\delta[{\omega'}^{a_{2}}_{a_{1}}(g^{a_{1}}-{f_{1}}^{a_{1}}+{Z^{(2)}}{}^{a_{1}}_{b_{2}}{g_{1}}^{b_{2}})]
\end{equation}
functional unities
\begin{align}
    \begin{gathered}
        1=\text{Det}(\omega'{}_{a_{1}}^{a_{2}}{Z^{(2)}}{}{}_{b_{2}}^{a_{1}})\int \mathscr{D}f_{2}\;\delta[\omega'{}^{a_{2}}_{a_{1}}(f_{1}{}^{a_{1}}+{Z^{(2)}}{}{}^{a_{1}}_{b_{2}}f_{2}{}^{b_{2}})]
    \end{gathered}\\
    \begin{gathered}
        1=\text{Det}({{\tilde{Z}^{(2)}}}{}_{a_{1}}^{b_{2}} \sigma'{}^{a_{1}}_{a_{2}})\int \mathscr{D}\xi\;\delta[(\pi_{a_{1}}+\xi_{b_{2}}{\tilde{Z}^{(2)}}{}_{a_{1}}^{b_{2}})\sigma'{}^{a_{1}}_{a_{2}}].
    \end{gathered}
\end{align}
After that, similarly to \eqref{eq:15}, \eqref{eq:18}, one can separate the integration over the third-stage gauge group and represent the modified delta function as the integral over extra ghosts, which leads to the expression of the form
\begin{align}
    \begin{gathered}
        \Delta_{m}^{-1}\;\text{Det}^{-1}(\omega'{}_{a_{1}}^{a_{2}}{Z^{(2)}}{}_{b_{2}}^{a_{1}})\; \text{Det}^{-1}({\tilde{Z}^{(2)}}{}^{b_{2}}_{a_{1}}\sigma'{}^{a_{1}}_{a_{2}} )
        = \\
        = \int \mathscr{D}f_{1}\mathscr{D}g_{1}\int\mathscr{D}\bar{c_{1}}\;\delta[\omega'{}^{a_{2}}_{a_{1}}f_{1}{}^{a_{1}}]\;\delta[\omega^{a_{1}}_{\alpha}(f^{\alpha}+{Z^{(1)}}{}^{\alpha}_{b_{1}}{f_{1}}^{b_{1}})+\sigma'{}^{a_{1}}_{a_{2}}\bar{c_{1}}^{a_{2}}]\;\delta[{\omega'}^{a_{2}}_{a_{1}}(g^{a_{1}}-{f_{1}}^{a_{1}}+{Z^{(2)}}{}^{a_{1}}_{b_{2}}{g_{1}}^{b_{2}})]
    \end{gathered}
\end{align}
Repeating the steps analogous to \eqref{eq:20}, and using the property \eqref{eq:21}, one obtains
\begin{align}
    \begin{gathered}
        \Delta_{m}^{-1}\;\text{Det}^{-1}(\omega'{}_{a_{1}}^{a_{2}}{Z^{(2)}}{}_{b_{2}}^{a_{1}})\; \text{Det}^{-1}({\tilde{Z}^{(2)}}{}_{a_{1}}^{b_{2}} \sigma'{}^{a_{1}}_{a_{2}})
        =\text{Det}^{-1}\begin{pmatrix}\omega Z^{(1)} & \sigma'{} & 0 \\ \omega' & 0 & 0 \\ -\omega' & 0 & \omega'{Z^{(2)}}\end{pmatrix}=\\=\text{Det}^{-1}(F')\;\text{Det}^{-1}(\omega'{}_{a_{1}}^{a_{2}}{Z^{(2)}}{}_{b_{2}}^{a_{1}}),
    \end{gathered}
\end{align}
where
\begin{equation}
F'{}^{a_{1}}_{b_{1}}=\omega^{a_{1}}_{\alpha}{Z^{(1)}}{}^{\alpha}_{b_{1}}+\sigma'{}^{a_{1}}_{a_{2}}\omega'^{a_{2}}_{b_{1}}.
\end{equation}
Finally one has
\begin{equation}
\Delta_{m}=\frac{\text{Det}(F')}{\text{Det}( {\tilde{Z}^{(2)}}{}_{a_{1}}^{b_{2}} \sigma'{}^{a_{1}}_{a_{2}} ) },
\end{equation}
and the modified measure and the modified delta function for second-stage gauge parameters both take the form:
\begin{align}
    \begin{gathered}
        \mathscr{D}\mu_{{f_{1},g_{1}}}=\mathscr{D}f_{1}\mathscr{D}g_{1}\;\text{Det}(\omega'{}_{a_{1}}^{a_{2}}{Z^{(2)}}{}_{b_{2}}^{a_{1}})\;
        \delta[\omega'{}^{a_{2}}_{a_{1}}f_{1}{}^{a_{1}}],\;\;\;\;\;\;\;\;\;\;\;\;\;\;\;\;\;\;\;\;\;\;\;\;\;\;\;
    \end{gathered}\\
    \begin{gathered}
        \hat{\delta}[\omega^{a_{1}}_{\alpha}(f^{\alpha}+{Z^{(1)}}{}^{\alpha}_{b_{1}}{f_{1}}^{b_{1}}))]
        =
        \text{Det}({\tilde{Z}^{(2)}}{}_{a_{1}}^{b_{2}}\sigma'{}^{a_{1}}_{a_{2}}) \int\mathscr{D}\bar{c_{1}}\;
        \delta[\omega^{a_{1}}_{\alpha}(f^{\alpha}+{Z^{(1)}}{}^{\alpha}_{b_{1}}{f_{1}}^{b_{1}})+\sigma'{}^{a_{1}}_{a_{2}}\bar{c_{1}}^{a_{2}}]
    \end{gathered}
\end{align}

Now we substitute the fully defined functional unity \eqref{eq:29} into the integral \eqref{eq:27} and perform the sequence of transformations analogous to \eqref{eq:18} (denoting the integrand in \eqref{eq:27} by $\Phi[\phi^{f},s^{f,g}]$):
\begin{align}
    \begin{gathered}
        \label{eq:40}
        \int\mathscr{D}f\mathscr{D}g\;\Phi[\phi^{f},s^{f,g}]\rightarrow\text{Det}(F')\int\mathscr{D}f\mathscr{D}g\;\Phi[\phi^{f},s^{f,g}]\int\mathscr{D}\mu_{{f_{1},g_{1}}}\int\mathscr{D}\bar{c_{1}}
\times
\\
\times
\delta[\omega^{a_{1}}_{\alpha}(f^{\alpha}+{Z^{(1)}}{}^{\alpha}_{b_{1}}{f_{1}}^{b_{1}})+\sigma'{}^{a_{1}}_{a_{2}}\bar{c_{1}}^{a_{2}}]\;
\delta[{\omega'}^{a_{2}}_{a_{1}}(g^{a_{1}}-{f_{1}}^{a_{1}}+{Z^{(2)}}{}^{a_{1}}_{b_{2}}{g_{1}}^{b_{2}})]=\\
        =V\times\text{Det}(F')\int\mathscr{D}f\mathscr{D}g\mathscr{D}\bar{c_{1}}\;\Phi[\phi^{f},s^{f,g}]\;
        \delta[\omega^{a_{1}}_{\alpha}f^{\alpha}+\sigma'{}^{a_{1}}_{a_{2}}\bar{c_{1}}^{a_{2}}]\;\delta[{\omega'}^{a_{2}}_{a_{1}}g^{a_{1}}].
    \end{gathered}
\end{align}
As a result, one obtains the expression for the modified integration measure over first-stage gauge parameters,
\begin{align}
    \begin{gathered}
        \label{eq:41}
        \mathscr{D}\mu_{f,g}=\mathscr{D}f\mathscr{D}g\;\text{Det}(F')\int\mathscr{D}\bar{c_{1}}\;
        \delta[\omega^{a_{1}}_{\alpha}f^{\alpha}+\sigma'{}^{a_{1}}_{a_{2}}\bar{c_{1}}^{a_{2}}]\;\delta[\omega'{}^{a_{2}}_{a_{1}}g^{a_{1}}].
    \end{gathered}
\end{align}

\subsection{Modified delta functions}

Now we define modified delta functions in the expression \eqref{eq:26}. The expressions for ordinary delta functions are
 \begin{align}
    \begin{gathered}
        \label{eq:42}
        \delta[\chi^{\alpha}(\phi^{i})]=\int \mathscr{D}\pi\;e^{i\pi_{\alpha}\chi^{\alpha}(\phi^{i})},
    \end{gathered}\\
    \begin{gathered}
        \label{eq:43}
        \delta[\omega^{a_{1}}_{\alpha}s^{\alpha}]=\int \mathscr{D}\theta\;e^{i\theta_{a_{1}}\omega^{a_{1}}_{\alpha}s^{\alpha}}.
    \end{gathered}
 \end{align}
It is easy to see that their exponentials are invariant under the transformations $\delta\pi_{\alpha} = \xi_{a_{1}} {\bar{Z}^{(1)}}{}_{\alpha}^{a_{1}}$, $\delta\theta_{a_{1}}=\eta_{a_{2}} {\tilde{Z}^{(2)}}{}_{a_{1}}^{a_{2}}$.
We fix their symmetries by the conditions $\sigma(\pi)=\pi_{\alpha}\sigma_{a_{1}}^{\alpha}$ and $\sigma'{}(\theta)=\theta_{a_{1}}\sigma'{}_{a_{2}}^{a_{1}}$. The matrix $\bar{Z}^{(1)}$, similarly to $Z^{(1)}$, possesses $m_{2}$ independent null vectors. This means that the expression $\xi_{a_{1}} {\bar{Z}^{(1)}}{}_{\alpha}^{a_{1}}$ itself is invariant under the transformation $\delta\xi_{a_{1}}=\xi_{1}{}_{a_{2}} \bar{Z}^{(2)}{}_{a_{1}}^{a_{2}}$, where $\bar{Z}^{(2)}$ denotes null vectors of the matrix $\bar{Z}^{(1)}$, while the expression for the delta function
 \begin{equation}
  \delta[\pi_{\alpha}\sigma_{a_{1}}^{\alpha}]
  = \int \mathscr{D}\pi_{1}\; e^{i \pi_{\alpha}\sigma_{a_{1}}^{\alpha}\pi_{1}{}^{a_{1}}}
 \end{equation}
is invariant under the transformation $\delta\pi_{1}{}^{a_{1}}=\tilde{\bar{Z}}^{(2)}{}^{a_{1}}_{a_{2}}\lambda^{a_{2}}$, where $\tilde{\bar{Z}}^{(2)}{}$ denotes left null vectors of gauge condition $\sigma$ (ranks of matrices $\bar{Z}^{(2)}$ and $\tilde{\bar{Z}}^{(2)}{}$ are equal to $m_{2}$).
Therefore, it is necessary to insert unities into expressions \eqref{eq:42} and \eqref{eq:43},
\begin{align}
    \begin{gathered}
        \label{eq:49}
        1=\Delta_{d}\int \mathscr{D}\mu_{\xi}\;\hat{\delta}[(\pi_{\alpha}+\xi_{b_{1}}{\bar{Z}^{(1)}}{}_{\alpha}^{b_{1}}) \sigma_{a_{1}}^{\alpha}],\;\;\;\;\;\;\;\;
    \end{gathered}\\
    \begin{gathered}
        \label{eq:50}
        1=\text{Det}({\tilde{Z}^{(2)}}{}_{a_{1}}^{b_{2}} \sigma'{}_{a_{2}}^{a_{1}}) \int \mathscr{D}\eta\; \delta[(\theta_{a_{1}}\!+\eta_{b_{2}}{\tilde{Z}^{(2)}}{}_{a_{1}}^{b_{2}})\sigma'{}_{a_{2}}^{a_{1}}].
    \end{gathered}
\end{align}

The measure $\mathscr{D}\mu_{\xi}$ and the delta function $\hat{\delta}[\pi_{\alpha}\sigma_{a_{1}}^{\alpha}]$ are defined similarly to the construction described above:
\begin{align}
    \begin{gathered}
        \mathscr{D}\mu_{\xi} =\mathscr{D}\xi\; \text{Det}(\bar{Z}^{(2)}{}_{a_{1}}^{b_{2}} \bar{\omega}_{a_{2}}^{a_{1}})\; \delta[\xi_{a_{1}}\bar{\omega}_{a_{2}}^{a_{1}}]\;\;\;\;\;\;\;\;\;\;
    \end{gathered}\\
    \begin{gathered}
        \hat{\delta}[\pi_{\alpha} \sigma_{a_{1}}^{\alpha}] =\text{Det}(\bar{\sigma}^{a_{2}}_{a_{1}}\tilde{\bar{Z}}^{(2)}{}^{a_{1}}_{b_{2}}) \int \mathscr{D}c' \,\delta[\pi_{\alpha}\sigma_{a_{1}}^{\alpha}+c'{}_{a_{2}}\bar{\sigma}_{a_{1}}^{a_{2}}],
    \end{gathered}
\end{align}
where $\bar{\omega}$ is gauge fixing the symmetry $\delta\xi_{a_{1}}=\xi_{1}{}_{a_{2}}\bar{Z}^{(2)}{}_{a_{1}}^{a_{2}}$ and $\bar{\sigma}$ is gauge fixing the symmetry $\delta\pi_{1}{}^{a_{1}}=\tilde{\bar{Z}}^{(2)}{}^{a_{1}}_{a_{2}}\lambda^{a_{2}}$, while the fields $c'$ play the role of second-stage extra ghosts. Substituting these expressions into \eqref{eq:49}, and proceeding in analogy with the previous steps, one obtains the expression for the determinant $\Delta_{d}$
\begin{align}
    \begin{gathered}
        1=\Delta_{d}\;\text{Det}(\bar{Z}^{(2)}{}_{a_{1}}^{b_{2}} \bar{\omega}_{a_{2}}^{a_{1}})\; \text{Det}(\bar{\sigma}^{a_{2}}_{a_{1}}\tilde{\bar{Z}}^{(2)}{}^{a_{1}}_{b_{2}})
        \int \mathscr{D}\xi\mathscr{D}c'\; \delta[\xi_{a_{1}}\bar{\omega}_{a_{2}}^{a_{1}}]\; \delta[(\pi_{\alpha}+\xi_{b_{1}}{\bar{Z}^{(1)}}{}_{\alpha}^{b_{1}})\sigma_{a_{1}}^{\alpha} + c'{}_{a_{2}}\bar{\sigma}_{a_{1}}^{a_{2}}]
\\
        \Delta_{d}=\frac{\text{Det}(F'')}{\text{Det}(\bar{Z}^{(2)}{}_{a_{1}}^{b_{2}} \bar{\omega}_{a_{2}}^{a_{1}})\; \text{Det}(\bar{\sigma}^{a_{2}}_{a_{1}}\tilde{\bar{Z}}^{(2)}{}^{a_{1}}_{b_{2}})},
        \;\;\;\;\;\;\;\;\;\;
        F''{}_{a_{1}}^{b_{1}} = {\bar{Z}^{(1)}}{}_{\alpha}^{b_{1}}\sigma_{a_{1}}^{\alpha} +\bar{\omega}_{a_{2}}^{b_{1}} \bar{\sigma}_{a_{1}}^{a_{2}}
    \end{gathered}
\end{align}

Now we substitute the fully defined unity \eqref{eq:49} into \eqref{eq:42} and perform the following sequence of transformations:
\begin{align}
    \begin{gathered}
        \label{eq:54}
        \delta[\chi^{\alpha}(\phi^{i})]=\int \mathscr{D}\pi\;e^{i\pi_{\alpha}\chi^{\alpha}(\phi^{i})}
\rightarrow
\\
\rightarrow
\int\mathscr{D}\mu_{\xi} \int \mathscr{D}\pi\mathscr{D}c'\; \frac{\text{Det}(F'')}{\text{Det}(\bar{Z}^{(2)}{}_{a_{1}}^{b_{2}} \bar{\omega}_{a_{2}}^{a_{1}})} \delta[(\pi_{\alpha}+\xi_{b_{1}}{\bar{Z}^{(1)}}{}_{\alpha}^{b_{1}})\sigma_{a_{1}}^{\alpha} +c'{}_{a_{2}}\bar{\sigma}_{a_{1}}^{a_{2}}]\;e^{i\pi_{\alpha}\chi^{\alpha}(\phi^{i})}
=
\\
=
\int\mathscr{D}\mu_{\xi}\,\frac{\text{Det}(F'')}{\text{Det}(\bar{Z}^{(2)}{}_{a_{1}}^{b_{2}} \bar{\omega}_{a_{2}}^{a_{1}})}\int \mathscr{D}\pi\mathscr{D}c\mathscr{D}c'\;e^{i\pi_{\alpha}\sigma_{a_{1}}^{\alpha}c^{a_{1}} +i c'{}_{a_{2}}\bar{\sigma}_{a_{1}}^{a_{2}}c^{a_{1}} +i\pi_{\alpha}\chi^{\alpha}(\phi^{i})}
=
\\
=
\left(\int\mathscr{D}\mu_{\xi}\right) \frac{\text{Det}(F'')}{\text{Det}(\bar{Z}^{(2)}{}_{a_{1}}^{b_{2}}\bar{\omega}_{a_{2}}^{a_{1}})} \int\mathscr{D}c\;\delta[\chi^{\alpha}(\phi^{i})+\sigma^{\alpha}_{a_{1}}c^{a_{1}}]\;\delta[\bar{\sigma}^{a_{2}}_{a_{1}}c^{a_{1}}].
    \end{gathered}
\end{align}
Here the integration over the second-stage gauge group has been disentangled, the delta function has been represented as an integral over first-stage extra ghosts, and the integration over second-stage extra ghosts has been carried out. Also we substitute the unity \eqref{eq:50} into the expression \eqref{eq:43} and perform the transformations analogous to \eqref{eq:15}. As a result, one obtains expressions for modified delta functions
\begin{align}
    \begin{gathered}
        \label{eq:55}
        \hat{\hat{\delta}}[\chi^{\alpha}(\phi^{i}+R^{i}_{\alpha}f^{\alpha})] =\frac{\text{Det}(F'')}{\text{Det}(
        \bar{Z}^{(2)}{}_{a_{1}}^{b_{2}}\bar{\omega}_{a_{2}}^{a_{1}})} \int\mathscr{D}c\;\delta[\chi^{\alpha}(\phi^{i}+R^{i}_{\alpha}f^{\alpha})
        +\sigma^{\alpha}_{a_{1}}c^{a_{1}}]\;\delta[\bar{\sigma}^{a_{2}}_{a_{1}}c^{a_{1}}],\qquad
    \end{gathered}\\
    \begin{gathered}
        \label{eq:56}
        \hat{\delta}[\omega^{a_{1}}_{\alpha}(s^{\alpha}+f^{\alpha}+{Z^{(1)}}{}^{\alpha}_{b_{1}}g^{b_{1}})]
        =
        \text{Det}({\tilde{Z}^{(2)}}{}_{a_{1}}^{b_{2}}\sigma'{}_{a_{2}}^{a_{1}})\int\mathscr{D}\tilde{c_{1}}\;
        \delta[\omega^{a_{1}}_{\alpha}(s^{\alpha}+f^{\alpha}+{Z^{(1)}}{}^{\alpha}_{b_{1}}g^{b_{1}})
        +\sigma'{}^{a_{1}}_{a_{2}}\tilde{c_{1}}^{a_{2}}].
    \end{gathered}
\end{align}

\begin{figure}
\centering
\includegraphics{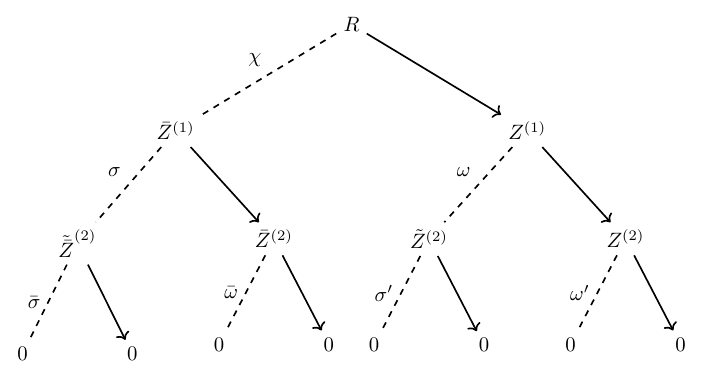}
\caption{Diagram of reducibility}
\label{fig:placeholder}
\end{figure}
Relationships between gauge generators and gauge fixing conditions are shown in Figure 1. The following notation is used: a node represents matrix of generators of gauge transformation at corresponding level. An arrow goes from a generator matrix to the matrix of its right zero vectors. A dashed line goes from a generator matrix to the matrix of left zero vectors of gauge fixing condition used to fix the symmetry associated with those generators (the gauge condition is indicated above the dashed line).

\subsection{Faddeev-Popov determinant and the functional integral}
In order to compute the Faddeev-Popov determinant we substitute expressions \eqref{eq:41}, \eqref{eq:55} and \eqref{eq:56} into the integral \eqref{eq:26} and perform the transformation analogous to \eqref{eq:20}:
\begin{align}
    \begin{gathered}
        \Delta^{-1}\;\frac{\text{Det}(\bar{Z}^{(2)}{}_{a_{1}}^{b_{2}} \bar{\omega}_{a_{2}}^{a_{1}})}{\text{Det}(F'')\;\text{Det}(F')\;
        \text{Det}({\tilde{Z}^{(2)}}{}_{a_{1}}^{b_{2}} \sigma'{}_{a_{2}}^{a_{1}})}
  =
  \\
  =
  \int\mathscr{D}f\mathscr{D}g\mathscr{D}\bar{c_{1}}\mathscr{D}c\mathscr{D}\tilde{c_{1}}\,\delta[\omega^{a_{1}}_{\alpha}f^{\alpha}
  +\sigma'{}^{a_{1}}_{a_{2}}\bar{c_{1}}^{a_{2}}]\;\delta[\omega'{}^{a_{2}}_{a_{1}}g^{a_{1}}]\;\delta[\chi^{\alpha}(\phi^{i}
  +R^{i}_{\alpha}f^{\alpha})+\sigma^{\alpha}_{a_{1}}c^{a_{1}}]\times\\
        \times\delta[\bar{\sigma}^{a_{2}}_{a_{1}}c^{a_{1}}]\;\delta[\omega^{a_{1}}_{\alpha}(s^{\alpha}+f^{\alpha}
        +{Z^{(1)}}{}^{\alpha}_{b_{1}}g^{b_{1}})+\sigma'{}^{a_{1}}_{a_{2}}\tilde{c_{1}}^{a_{2}}]=\\
        =\text{Det}^{-1}\begin{pmatrix}\frac{\delta\chi}{\delta\phi}R & \sigma & 0 & 0 & 0\\ \omega & 0 & \sigma'{} & 0 & 0 \\ 0 & \bar{\sigma} & 0 & 0 & 0 \\ \omega & 0 & 0 & \omega Z^{(1)} & \sigma'{} \\ 0 & 0 & 0 & \omega' & 0\end{pmatrix} = \text{Det}^{-1}\begin{pmatrix}\frac{\delta\chi}{\delta\phi}R & \sigma & 0\\ \omega & 0 & \sigma'{}\\ 0 & \bar{\sigma} & 0\end{pmatrix}\;\text{Det}^{-1}(F')\equiv\text{Det}^{-1}(H)\text{Det}^{-1}(F').
    \end{gathered}
\end{align}

The determinant of the newly introduced matrix $H$ can be transformed as follows. Consider the operator
\begin{equation}
G=\begin{pmatrix}\frac{\delta\chi}{\delta\phi}R & \sigma\\ \omega & 0\end{pmatrix}+\begin{pmatrix}0 \\ \sigma'{}\end{pmatrix}\begin{pmatrix}0 & \bar{\sigma}\end{pmatrix}=\begin{pmatrix}\frac{\delta\chi}{\delta\phi}R & \sigma\\ \omega & \sigma'{}\bar{\sigma}\end{pmatrix}.
\end{equation}
It is straightforward to verify that it satisfies the properties,
\begin{align}
    \begin{gathered}
        G\begin{pmatrix}0 \\ \tilde{\bar{Z}}^{(2)}{}\end{pmatrix}=\begin{pmatrix}0 \\ \sigma'{}\end{pmatrix}\begin{pmatrix}\bar{\sigma}\tilde{\bar{Z}}^{(2)}{}\end{pmatrix},\qquad
        \begin{pmatrix}0 & {\tilde{Z}^{(2)}}\end{pmatrix}G=\begin{pmatrix}{\tilde{Z}^{(2)}}\sigma'{}\end{pmatrix}\begin{pmatrix}0 & \bar{\sigma}\end{pmatrix},
    \end{gathered}
\end{align}
which immediately implies that
\begin{equation}
\label{eq:62}
\begin{pmatrix}0 & \bar{\sigma}\end{pmatrix}\;G^{-1}\begin{pmatrix}0 \\ \sigma'{}\end{pmatrix}=I.
\end{equation}
Therefore, multiplying the matrix $H$ by a matrix with unit determinant, constructed on the basis of relation \eqref{eq:62}, one obtains
\begin{equation}
\text{Det}(H)=\text{Det}\left[\begin{pmatrix}\frac{\delta\chi}{\delta\phi}R & \sigma & 0\\ \omega & 0 & \sigma'{}\\ 0 & \bar{\sigma} & 0\end{pmatrix}\begin{pmatrix}\begin{pmatrix}I & 0 \\ 0 & I\end{pmatrix} & G^{-1}\!\begin{pmatrix}0 \\ \sigma'{}\end{pmatrix} \\[1.7ex]
\begin{pmatrix}0 & \bar{\sigma}\end{pmatrix} & 0 \end{pmatrix}\right]
=\text{Det}\begin{pmatrix}\frac{\delta\chi}{\delta\phi}R & \sigma & 0 \\ \omega & \sigma'{}\bar{\sigma} & 0 \\ 0 & \bar{\sigma} & I\end{pmatrix}=\text{Det}(G).
\end{equation}
An analogous procedure can be carried out for the operator $G$. Consider the operator gauge-fixed by $\omega$ and $\sigma$ matrices,
\begin{equation}
\bar{F}^{\alpha}_{\beta}=\frac{\delta\chi^{\alpha}}{\delta\phi^{i}}R^{i}_{\beta}+\sigma^{\alpha}_{a_{1}}\omega^{a_{1}}_{\beta}.
\end{equation}
By construction, it satisfies the properties $\bar{Z}^{(1)}\bar{F}=\bar{Z}^{(1)}\sigma\omega$, $\bar{F}Z^{(1)}=\sigma\omega Z^{(1)}$,
which lead to the relation $\omega \bar{F}^{-1}\sigma=I$. It is then easy to see that
\begin{equation}
\text{Det}(G)=\text{Det}\left[\begin{pmatrix}\frac{\delta\chi}{\delta\phi}R & \sigma\\ \omega & \sigma'{}\bar{\sigma}\end{pmatrix}\begin{pmatrix}I & \bar{F}^{-1}\sigma \\ \omega & 0\end{pmatrix}\right]=\text{Det}\begin{pmatrix} \bar{F} & 0\\ \omega+\sigma'{}\bar{\sigma}\omega & I\end{pmatrix}=\text{Det}(\bar{F}),
\end{equation}
where the matrix $G$ has been multiplied by a matrix with unit determinant. As a result, one obtains the expression for the Faddeev–Popov determinant
\begin{equation}
 \Delta =\frac{\text{Det}(\bar{Z}^{(2)}{}_{a_{1}}^{b_{2}}\bar{\omega}_{a_{2}}^{a_{1}})\text{Det}(\bar{F})} {\text{Det}(F'')\;\text{Det}({\tilde{Z}^{(2)}}{}_{a_{1}}^{b_{2}}\sigma'{}_{a_{2}}^{a_{1}})}.
\end{equation}
Further, performing the sequence of transformations analogous to \eqref{eq:24} and \eqref{eq:24},
\begin{align}
    \begin{gathered}
         \int \mathscr{D}\phi\mathscr{D}s\;e^{iS[\phi,s]}\rightarrow\\\rightarrow\left(\int\mathscr{D}\mu_{f,g}\right)\;\text{Det}(\bar{F})\int\mathscr{D}\phi\mathscr{D}s\mathscr{D}\bar{c_{1}}\mathscr{D}c\mathscr{D}\tilde{c_{1}}\;\delta[\chi^{\alpha}(\phi^{i})+\sigma^{\alpha}_{a_{1}}c^{a_{1}}]\;\delta[\bar{\sigma}^{a_{2}}_{a_{1}}c^{a_{1}}]\;\delta[\omega^{a_{1}}_{\alpha}s^{\alpha}+\sigma'{}^{a_{1}}_{a_{2}}\tilde{c_{1}}^{a_{2}}]\;e^{iS[\phi,s]},
    \end{gathered}
\end{align}
one arrives at the final expression for the functional integral of a massive second-stage reducible gauge theory with Stueckelberg fields
\begin{equation}
Z=\text{Det}(\bar{F})\int\mathscr{D}\phi\mathscr{D}s\mathscr{D}\bar{c_{1}}\mathscr{D}c\mathscr{D}\tilde{c_{1}}\;
\delta[\chi^{\alpha}(\phi^{i})+\sigma^{\alpha}_{a_{1}}c^{a_{1}}]\;\delta[\bar{\sigma}^{a_{2}}_{a_{1}}c^{a_{1}}]\;
\delta[\omega^{a_{1}}_{\alpha}s^{\alpha}+\sigma'{}^{a_{1}}_{a_{2}}\tilde{c_{1}}^{a_{2}}]\;e^{iS[\phi,s]}.
\end{equation}
Note that in the subsequent computation of the effective action a complication may arise due to the presence of nondiagonal terms in the action $S[\phi,s]$ mixing the fields $\phi$ and $s$. However, recall that chosen gauge conditions are not the only possible ones. At each stage of the quantization, gauge invariance of corresponding stage ensures independence of the result from a particular gauge choice. In order to eliminate mixed terms in the action, one should impose gauge conditions fixing appropriate linear combinations of fields $\phi$ and $s$. A specific choice of such linear combinations depends on model under consideration.

\section{Quantization of massive totally antisymmetric tensor spinor theory}
 \label{Sect:spin-tensor}
\subsection{Quantization of the first-stage reducible theory with mass deformation}

We consider the model of rank 2 massive totally antisymmetric tensor spinor field in $AdS_{d}$ space. The model is described by the action (we use the notation as in \cite{BBKN2})
\begin{align}
    \begin{gathered}
        \label{eq:69}
        S[\psi_{\mu\nu},\bar{\psi}_{\mu\nu}]
        =i\!\int\! d^{d}x\;\sqrt{|g|}\;\bar{\psi}_{\mu_{1}\mu_{2}}\gamma^{\mu_{1}\mu_{2}\sigma\nu_{1}\nu_{2}}D_{\sigma}\psi_{\nu_{1}\nu_{2}}
        -m \!\int\! d^{d}x\;\sqrt{|g|}\;\bar{\psi}_{\mu_{1}\mu_{2}}\gamma^{\mu_{1}\mu_{2}\nu_{1}\nu_{2}}\psi_{\nu_{1}\nu_{2}},\\
         D_{\mu}=\nabla_{\mu}\pm i\frac{\sqrt{r}}{2}\gamma_{\mu},
    \end{gathered}
\end{align}
where $\psi_{\mu\nu}=-\psi_{\nu\mu}$ is an antisymmetric tensor spinor, $\bar{\psi}_{\mu\nu}$ is the Dirac-conjugated tensor spinor, $\nabla_{\mu}$ is a covariant derivative, and $r$ is  determined from $R^{\mu\nu}{}_{\rho\sigma}=r(\delta^\mu_\sigma\delta^\nu_\rho-\delta^\mu_\rho\delta^\nu_\sigma)$ and $r>0$ for the $AdS$ space. In order for the product $\gamma^{\mu_{1}\mu_{2}\sigma\nu_{1}\nu_{2}}$ to be well defined, the spacetime dimension $d$ must satisfy the lower bound condition $d>4$. This inequality is always fulfilled in this subsection. Operator $D_{\mu}$ satisfies the property $D_{[\mu_{1}}D_{\mu_{2}}f_{\mu_{3}...\mu_{n}]}=0$, where $f_{\mu_{3}...\mu_{n}}$ is an arbitrary antisymmetric tensor spinor. As a consequence, the kinetic term of the action \eqref{eq:69} is invariant under gauge transformations $\delta\psi_{\mu\nu}=D_{[\mu}f_{\nu]}$. In order to restore gauge invariance of the mass term, we introduce Stueckelberg fields $\lambda_{\mu}$, $\bar{\lambda}_{\mu}$:
\begin{align}
    \begin{gathered}
         S[\psi,\bar{\psi},\lambda,\bar{\lambda}]=i\int d^{d}x\;\sqrt{|g|}\;\bar{\psi}_{\mu_{1}\mu_{2}}\gamma^{\mu_{1}\mu_{2}\sigma\nu_{1}\nu_{2}}D_{\sigma}\psi_{\nu_{1}\nu_{2}}
-
\\
-
m\int d^{d}x\;\sqrt{|g|}\;\left(\bar{\psi}_{\mu_{1}\mu_{2}}+\frac{1}{m}\bar{\lambda}_{[\mu_{1}}\overleftarrow{D}_{\mu_{2}]}\right)
         \gamma^{\mu_{1}\mu_{2}\nu_{1}\nu_{2}}\left(\psi_{\nu_{1}\nu_{2}}-\frac{1}{m}D_{[\nu_{1}}\lambda_{\nu_{2}]}\right).
    \end{gathered}
\end{align}
The operator $\overleftarrow{D}_{\mu}$ here is defined as $\bar{f}\overleftarrow{D}_{\mu}=\bar{f}\overleftarrow{\nabla}_{\mu}\mp i(1/2)\sqrt{r}\bar{f}\gamma_{\mu}$ (the covariant derivative acts to the left). The resulting action is invariant under the transformation
\begin{equation}
\label{eq:72}
\delta\psi_{\mu\nu}=D_{[\mu}f_{\nu]},\qquad \delta\lambda_{\mu}=mf_{\mu}+D_{\mu}g
\end{equation}
and can be brought to the form
\begin{align}
    \begin{gathered}
        \label{eq:73}
         S[\psi,\bar{\psi},\lambda,\bar{\lambda}]=S_{0}\mp i\frac{\sqrt{r}}{m}(d-3)S_{St}+S_{ND},
    \end{gathered}
\end{align}
where
\begin{itemize}
\item{$S_{0}$ denotes the action of the model prior to the introduction of Stueckelberg fields:}
$$S_{0}=i\int d^{d}x\;\sqrt{|g|}\;\bar{\psi}_{\mu_{1}\mu_{2}}\gamma^{\mu_{1}\mu_{2}\sigma\nu_{1}\nu_{2}}D_{\sigma}\psi_{\nu_{1}\nu_{2}}-m\int d^{d}x\;\sqrt{|g|}\;\bar{\psi}_{\mu_{1}\mu_{2}}\gamma^{\mu_{1}\mu_{2}\nu_{1}\nu_{2}}\psi_{\nu_{1}\nu_{2}}$$
\item{$S_{St}$ is the diagonal contribution of Stueckelberg fields:}
$$S_{St}=\int d^{d}x\;\sqrt{|g|}\;\bar{\lambda}_{\mu}\gamma^{\mu\sigma\nu}D_{\sigma}\lambda_{\nu}$$
\item{$S_{ND}$ represents the non-diagonal contributions:}
$$
S_{ND}=-\int d^{d}x\;\sqrt{|g|}\;\bar{\lambda}_{\mu_{1}}\overleftarrow{D}_{\mu_{2}}\gamma^{\mu_{1}\mu_{2}\nu_{1}\nu_{2}}\psi_{\nu_{1}\nu_{2}}+\int d^{d}x\;\sqrt{|g|}\;\bar{\psi}_{\mu_{1}\mu_{2}}\gamma^{\mu_{1}\mu_{2}\nu_{1}\nu_{2}}D_{\nu_{1}}\lambda_{\nu_{2}}
$$
\end{itemize}
Here and below, the following property of antisymmetrized products of gamma matrices is used
\begin{equation}
\gamma_\sigma\gamma^{\sigma\mu_1\ldots\mu_m}
=
\gamma^{\mu_1\ldots\mu_m\sigma}\gamma_{\sigma}
=
(d-m)\gamma^{\mu_1\ldots\mu_m}
\end{equation}

The action \eqref{eq:73} contains nondiagonal terms mixing fields $\psi_{\mu\nu}$ and $\lambda_{\mu}$ (the last two terms). Diagonalization of the action is performed in two steps. First, one can eliminate derivative cross terms by means of the linear field redefinition:
\begin{equation}
\label{eq:75}
\psi_{\mu\nu}=\phi_{\mu\nu}-\frac{i}{d-4}\gamma_{[\mu}\lambda_{\nu]}
\end{equation}
After this redefinition, the action takes the form
\begin{align}
    \begin{gathered}
        \label{eq:76}
        S[\phi,\bar{\phi},\lambda,\bar{\lambda}]=\tilde{S}_{0}-\frac{p^{(2)}}{m}\tilde{S}_{St}+p^{(2)}\tilde{S}_{ND},\qquad
        p^{(2)}\equiv(d-3)\left(\pm\sqrt{r}-\frac{m}{d-4}\right),
    \end{gathered}
\end{align}
where
\begin{itemize}
\item{$\tilde{S}_{0}$ is the diagonal contribution of the field $\phi$:}
$$\tilde{S}_{0}=i\int d^{d}x\;\sqrt{|g|}\;\bar{\phi}_{\mu_{1}\mu_{2}}\gamma^{\mu_{1}\mu_{2}\sigma\nu_{1}\nu_{2}}D_{\sigma}\phi_{\nu_{1}\nu_{2}}-m\int d^{d}x\;\sqrt{|g|}\;\bar{\phi}_{\mu_{1}\mu_{2}}\gamma^{\mu_{1}\mu_{2}\nu_{1}\nu_{2}}\phi_{\nu_{1}\nu_{2}},$$
\item{$\tilde{S}_{St}$ is the diagonal contribution of Stueckelberg fields including the additional term arising after the substitution \eqref{eq:75}:}
$$\tilde{S}_{St}=i\int d^{d}x\;\sqrt{|g|}\;\bar{\lambda}_{\mu}\gamma^{\mu\nu\rho}D_{\nu}\lambda_{\rho}+m\frac{d-2}{d-4}\int d^{d}x\;\sqrt{|g|}\;\bar{\lambda}_{\mu}\gamma^{\mu\nu}\lambda_{\nu},$$
\item{$\tilde{S}_{ND}$ denotes non-diagonal contributions after the substitution \eqref{eq:75}:}
$$\tilde{S}_{ND}=i\int d^{d}x\;\sqrt{|g|}\;\bar{\lambda}_{\mu}\gamma^{\mu\nu\rho}\phi_{\nu\rho}+i\int d^{d}x\;\sqrt{|g|}\;\bar{\phi}_{\mu\nu}\gamma^{\mu\nu\rho}\lambda_{\rho}.$$
\end{itemize}

One observes that, in addition to the disappearance of derivative cross terms, auxiliary fields $\lambda$ and $\bar\lambda$ acquire the mass and new non-derivative cross terms appear.

We now choose gauge conditions fixing the symmetry \eqref{eq:72}. The goal is to find such conditions which, upon integration with the corresponding delta functions, ensure cancellation of non-derivative cross terms. To this end, we employ the following decomposition of fields $\phi_{\mu\nu}$ and $\lambda_{\mu}$ into gamma-irreducible components
\begin{align}
    \begin{gathered}
        \label{eq:78}
        \phi_{\mu\nu}=\Phi_{\mu\nu}+\gamma_{[\mu}\Phi_{\nu]}+\gamma_{\mu\nu}\Phi,\qquad\gamma^{\mu}\Phi_{\mu\nu}=0,\;\gamma^{\mu}\Phi_{\mu}=0,\\
         \lambda_{\mu}=\Lambda_{\mu}+\gamma_{\mu}\Lambda,\qquad\gamma^{\mu}\Lambda_{\mu}=0.
    \end{gathered}
\end{align}
Consider the operator ${P_{1}}_{\mu}$ acting on antisymmetric fermionic rank-2 fields as follows:
\begin{equation}
\label{eq:79}
{P_{1}}_{\mu}(\lambda_{\alpha\beta})=\gamma^{\nu}\lambda_{\mu\nu}-\frac{1}{d}\gamma_{\mu}\gamma^{\alpha\beta}\lambda_{\alpha\beta}.
\end{equation}
Substituting the decomposition \eqref{eq:78} into \eqref{eq:79}, one readily obtains
\begin{equation}
{P_{1}}_{\mu}(\phi_{\alpha\beta})=\frac{1}{2}(d-2)\Phi_{\mu}.
\end{equation}
Thus, the operator ${P_{1}}_{\mu}$ projects from rank-2 field $\phi_{\mu\nu}$ its gamma-irreducible rank-1 component. We express non-derivative cross terms in terms of gamma-irreducible components of fields $\phi_{\mu\nu}$ and $\lambda_{\mu}$:
\begin{equation}
\mathscr{L}_{ND}\equiv\bar{\lambda}_{\nu}\gamma^{\nu\alpha\beta}\phi_{\alpha\beta}=\frac{(d-2)!}{(d-3)!}\bar{\Lambda}_{\mu}\Phi^{\mu}+\frac{d!}{(d-3)!}\bar{\Lambda}\Phi
\end{equation}
The conjugated counterterm $\bar{\mathscr{L}}_{ND}$ is obtained analogously. One can see that only the components of the same order are mixed. Therefore, if one chooses gauge conditions
\begin{equation}
{P_{1}}_{\mu}(\phi_{\alpha\beta})=\frac{1}{2}(d-2)\Phi_{\mu}=0,\qquad
\gamma^{\mu}\lambda_{\mu}=d\Lambda=0,
\end{equation}
which project out from the field $\phi_{\mu\nu}$ its first-order component and from field $\lambda_{\mu}$ the zeroth-order component (and corresponding Dirac-conjugated conditions for fields $\bar{\phi}_{\mu\nu}$, $\lambda_{\mu}$), then, after integration of the functional $F[\mathscr{L}_{ND},\bar{\mathscr{L}}_{ND}]$ with delta functions $\delta[\Phi_{\mu}];\delta[\bar{\Phi}_{\mu}];\delta[\Lambda];\delta[\bar{\Lambda}]$, the non-derivative cross terms vanish.

In terms of the original fields prior to their redefinition, the chosen gauge conditions take the form
\begin{equation}
{P_{1}}_{\mu}\left(\psi_{\alpha\beta}+\frac{i}{d-4}\gamma_{[\alpha}\lambda_{\beta]}\right)=0,\qquad
\gamma^{\mu}\lambda_{\mu}=0.
\end{equation}

We now proceed directly to quantization procedure. The symmetry \eqref{eq:72} is invariant under the transformation
\begin{equation}
\label{eq:83}
\delta f_{\mu}=D_{\mu}{f_{1}}\qquad \delta g=-m{f_{1}},
\end{equation}
and the operator ${P_{1}}_{\mu}$ introduced above satisfies the property $\gamma^{\mu}{P_{1}}_{\mu}=0$. Therefore, the model under consideration is a first-stage reducible gauge theory.

Following the scheme described in Sec. 3, we insert into the integral
\begin{equation}
\label{eq:84}
Z=\int\mathscr{D}\bar{\psi}\mathscr{D}\psi\mathscr{D}\bar{\lambda}\mathscr{D}\lambda\;e^{iS[\bar{\psi},\psi,\bar{\lambda},\lambda]}
\end{equation}
the functional unity constructed fout of modified delta function and modified integration measure (and the analogous unity for Dirac-conjugated fields).
\begin{align}
    \begin{gathered}
        \label{eq:85}
        1=\Delta^{-1}\int \mathscr{D}\mu_{f,g}\;\hat{\delta}\left[{P_{1}}_{\mu}\left(\phi_{\alpha\beta}+D_{[\alpha}f_{\beta]}+i\frac{m}{d-4}\gamma_{[\alpha}f_{\beta]}+i\frac{1}{d-4}\gamma_{[\alpha}D_{\beta]}g\right)\right]\times\\
        \times\delta\left[\gamma^{\mu}\left(\lambda_{\mu}+mf_{\mu}+D_{\mu}g\right)\right].
    \end{gathered}
\end{align}
Here the power of the determinant factor $\Delta$ is chosen negative in view of the Grassmann statistics of integration variables.

We now define the modified integration measure $\mathscr{D}\mu_{f,g}$. The integrand in \eqref{eq:85} is invariant under the transformation \eqref{eq:83}. We fix this symmetry by the condition
\begin{equation}
\gamma^{\mu}\left(f_{\mu}-i\frac{1}{d-4}\gamma_{\mu}g\right)=0
\end{equation}
and insert into the naively defined integral,
\begin{align}
    \begin{gathered}
        \int \mathscr{D}f\mathscr{D}g\;\delta\left[{P_{1}}_{\mu}\left(\phi_{\alpha\beta}
        +D_{[\alpha}f_{\beta]}+i\frac{m}{d-4}\gamma_{[\alpha}f_{\beta]}+i\frac{1}{d-4}\gamma_{[\alpha}D_{\beta]}g\right)\right]\times\\
        \times\delta\left[\gamma^{\mu}\left(\lambda_{\mu}+mf_{\mu}+D_{\mu}g\right)\right],
    \end{gathered}
\end{align}
the functional unity
\begin{align}
\begin{gathered}
1=\big(\Delta^{(2)}_{0}\big)^{-1}\int \mathscr{D}f_{1}\;\delta\left[\gamma^{\mu}\left(f_{\mu}-i\frac{1}{d-4}\gamma_{\mu}g+D_{\mu}f_{1}+i\frac{m}{d-4}\gamma_{\mu}f_{1}\right)\right],\\
\Delta_{p}^{(l)}=\text{Det}^{-1}\left(i\slashed{\nabla}_{p}-m\frac{d-2p}{d-2l}\right),
\end{gathered}
\end{align}
where $\slashed{\nabla}_{p}=\gamma^{\mu}\nabla_{\mu}\pm i(1/2)\sqrt{r}(d-2p)$. Performing the transformations analogous to \eqref{eq:18}, we obtain the expression for the modified integration measure
\begin{equation}
\label{eq:89}
\mathscr{D}\mu_{f,g}=\mathscr{D}f\mathscr{D}g\;\big(\Delta^{(2)}_{0}\big)^{-1}\;\delta\left[\gamma^{\mu}\left({f}_{\mu}-i\frac{1}{d-4}\gamma_{\mu}g\right)\right].
\end{equation}

Then we determine the modified delta function. The exponential in the expression
\begin{equation}
\label{eq:90}
\delta[{P_{1}}_{\mu}(\phi_{\alpha\beta})]=\int \mathscr{D}\bar{\pi}\;e^{i\bar{\pi}^{\mu}{P_{1}}_{\mu}(\phi_{\alpha\beta})}
\end{equation}
is invariant under the transformation $\delta\bar{\pi}_{\mu}=\bar{\xi}\gamma_{\mu}$. We fix this symmetry by imposing on $\bar{\pi}$ the condition $\bar{\pi}^{\mu}\gamma_{\mu}=0$ and insert into the expression \eqref{eq:90} the functional unity of the form
\begin{equation}
1=\int\mathscr{D}\bar{\xi}\;\delta[(\bar{\pi}_{\mu}+\bar{\xi}\gamma_{\mu})\gamma^{\mu}],
\end{equation}
where the determinant of the ultralocal operator $\gamma^{\mu}\gamma_{\mu}\delta(x,y)$ has been omitted. Performing transformations analogous to \eqref{eq:15}, we obtain the expression for the modified delta function
\begin{equation}
\label{eq:93}
\delta[{P_{1}}_{\mu}(\phi_{\alpha\beta})]=\int\mathscr{D}\bar{\pi}\;e^{i\bar{\pi}^{\mu}{P_{1}}_{\mu}(\phi_{\alpha\beta})}\;\delta[\bar{\pi}^{\mu}\gamma_{\mu}]
=\int\mathscr{D}c\;\delta[{P_{1}}_{\mu}(\phi_{\alpha\beta})+\gamma_{\mu}c].
\end{equation}
We now substitute the expressions \eqref{eq:89} and \eqref{eq:93} into the integral \eqref{eq:85}
\begin{align}
    \begin{gathered}
        \label{eq:94}
        \Delta=\big(\Delta^{(2)}_{0}\big)^{-1}\int \mathscr{D}f\mathscr{D}g\;\delta\left[\gamma^{\mu}\left(f_{\mu}-i\frac{1}{d-4}\gamma_{\mu}g\right)\right]\times\\\times
        \int\mathscr{D}c\;\delta\left[{P_{1}}_{\mu}\left(\phi_{\alpha\beta}+D_{[\alpha}{f}_{\beta]}+i\frac{m}{d-4}\gamma_{[\alpha}{f}_{\beta]}
        +i\frac{1}{d-4}\gamma_{[\alpha}D_{\beta]}g\right)+\gamma_{\mu}c\right]\times\\
        \times\delta\left[\gamma^{\mu}\left(\lambda_{\mu}+m\left({f}_{\mu}-i\frac{1}{d-4}
        \gamma_{\mu}g\right)+D_{\mu}g+i\frac{m}{d-4}\gamma_{\mu}g\right)\right].
    \end{gathered}
\end{align}
Note that, since the operator ${P_{1}}_{\mu}$ annihilates the zeroth gamma-irreducible component of the field it acts upon, the expression inside the second delta function does not depend on contributions of the form $\gamma_{\alpha\beta}g$. Using this property, we can write
\begin{align}
    \begin{gathered}
         {P_{1}}_{\mu}\left(\phi_{\alpha\beta}+D_{[\alpha}f_{\beta]}
         +i\frac{m}{d-4}\gamma_{[\alpha}f_{\beta]}+i\frac{1}{d-4}\gamma_{[\alpha}D_{\beta]}g\right)
         ={P_{1}}_{\mu}\left(\phi_{\alpha\beta}
         +D_{[\alpha}{F}_{\beta]}+i\frac{m}{d-4}\gamma_{[\alpha}{F}_{\beta]}\right),
    \end{gathered}
\end{align}
where
\begin{equation}
{F}_{\mu}={f}_{\mu}-i\frac{1}{d-4}\gamma_{\mu}g.   \label{eq:96}
\end{equation}
Then we substite \eqref{eq:96} into the integral \eqref{eq:94}:
\begin{align}
    \begin{gathered}
         \Delta=\big(\Delta^{(2)}_{0}\big)^{-1}\int \mathscr{D}F\mathscr{D}g\;\delta\left[\gamma^{\mu}{F}_{\mu}\right]\int\mathscr{D}c\;\delta\left[{P_{1}}_{\mu}\left(\phi_{\alpha\beta}
         +D_{[\alpha}{F}_{\beta]}+i\frac{m}{d-4}\gamma_{[\alpha}{F}_{\beta]}\right)+\gamma_{\mu}c\right]\times\\
         \times\delta\left[\gamma^{\mu}\left(\lambda_{\mu}+D_{\mu}g+i\frac{m}{d-4}\gamma_{\mu}g\right)\right].\label{eq:97}
    \end{gathered}
\end{align}
Here we also took into account that the contribution of $F_\mu$ in the last delta function vanishes due to $\delta\left[\gamma^{\mu}{F}_{\mu}\right]$.

One can represent delta functions in \eqref{eq:97} in the form of integrals over auxiliary variables $\bar\pi$ and $\bar\eta$,
\begin{align}
    \begin{gathered}
         \Delta=\big(\Delta^{(2)}_{0}\big)^{-1}\int \mathscr{D}F\mathscr{D}g\;\delta\left[\gamma^{\mu}{F}_{\mu}\right]\;\int\mathscr{D}\bar{\pi}\;\exp\left\{i\bar{\pi}^{\mu}{P_{1}}_{\mu}\left(\phi_{\alpha\beta}+D_{[\alpha}{F}_{\beta]}+i\frac{m}{d-4}\gamma_{[\alpha}{F}_{\beta]}\right)\right\}\;\delta[\bar{\pi}^{\mu}\gamma_{\mu}]\times\\
         \times\int\mathscr{D}\bar{\eta}\; \exp\left\{i\bar{\eta}\gamma^{\mu}\left(\lambda_{\mu}+D_{\mu}g+i\frac{m}{d-4}\gamma_{\mu}g\right)\right\},
    \end{gathered}
\end{align}
and decompose the variables $\bar{\pi}^{\mu}$, $F^{\mu}$ into gamma-irreducible components:
\begin{align}
    \begin{gathered}
         \Delta=\big(\Delta^{(2)}_{0}\big)^{-1}\int \mathscr{D}{\Gamma}_{\mu}\mathscr{D}
         \Gamma\mathscr{D}g\;\int\mathscr{D}\bar{\Pi}_{\mu}\mathscr{D}\bar{\Pi}\;
         \exp\left[i\bar{\Pi}^{\mu}\left\{{P_{1}}_{\mu}\left(\phi_{\alpha\beta}\right)
         -\slashed{\nabla}_{1}{\Gamma}_{\mu}-im\frac{d-2}{d-4}{\Gamma}_{\mu}\right\}+ \ldots \right]\times\\
         \times\delta[\Gamma]\delta[\bar{\Pi}]\int\mathscr{D}\bar{\eta}\;
         \exp\left[i\bar{\eta}\left(\gamma^{\mu}\lambda_{\mu}+\slashed\nabla_{0}g+im\frac{d}{d-4}g\right)\right].
    \end{gathered}
\end{align}
Here $\bar{\Pi}_{\mu},\bar{\Pi}$ are gamma-irreducible components of $\bar{\pi}_{\mu}$, while ${\Gamma}_{\mu},\Gamma$ are gamma-irreducible components of $F$. Ellipsis “...” denotes the terms containing components $\bar{\Pi}$ and $\Gamma$, which are annihilated by delta functions $\delta[\bar{\Pi}]$ and $\delta[\Gamma]$.
As a result, shifting the integration variables
\begin{align}
    \begin{gathered}
         \Gamma_{\mu}\rightarrow\Gamma'_{\mu}-\left(\slashed{\nabla}_{1}+im\frac{d-2}{d-4}\right)^{-1}{P_{1}}_{\mu}\left(\phi_{\alpha\beta}\right),\qquad
         g\rightarrow g'-\left(\slashed{\nabla}_{0}+im\frac{d}{d-4}\right)^{-1}\gamma^{\mu}\lambda_{\mu}
    \end{gathered}
\end{align}
and expressing Gaussian integrals over irreducible components in terms of the determinants of the corresponding operators, one can obtain the expression for the sought for Faddeev–Popov determinant,
\begin{equation}
\Delta=\Delta^{(2)}_{1}\times\frac{\Delta^{(2)}_{0}}{\Delta^{(2)}_{0}}
=\Delta^{(2)}_{1}.
\end{equation}

Substituting the fully defined functional unity \eqref{eq:85} (and the analogous unity for conjugated fields $\bar{\psi}_{\mu\nu}$, $\bar{\lambda}_{\mu}$) into the integral \eqref{eq:84}, making the change of integration variables $\psi_{\mu\nu}\rightarrow\phi_{\mu\nu}$ ($\phi_{\mu\nu}$ is defined by \eqref{eq:75}) and carrying out a sequence of transformations analogous to \eqref{eq:24}, we obtain
\begin{align}
    \begin{gathered}
         Z^{(2)}_{m}=\big(\Delta^{(2)}_{1}\big)^{-2}\int\mathscr{D}\bar{\phi}\mathscr{D}\phi\mathscr{D}\bar{\lambda}\mathscr{D}\lambda
         \mathscr{D}\bar{c}\mathscr{D}c\times\\
         \times e^{iS[\bar{\phi},\phi,\bar{\lambda},\lambda]}\;\delta[{P_{1}}_{\mu}(\phi_{\alpha\beta})+\gamma_{\mu}c]\;
         \delta[(\bar{\phi}_{\alpha\beta})\overleftarrow{{P_{1}}_{\mu}}+\bar{c}\gamma_{\mu}]\;
         \delta[\gamma^{\mu}\lambda_{\mu}]\;\delta[\bar{\lambda}^{\mu}\gamma_{\mu}],
    \end{gathered}
\end{align}
where $(\bar{\phi}_{\alpha\beta})\overleftarrow{{P_{1}}_{\mu}}\equiv\overline{{P_{1}}_{\mu}(\phi_{\alpha\beta})}$. Then we decompose integration variables into gamma-irreducible components
\begin{align}
    \begin{gathered}
         Z^{(2)}_{m}=\big(\Delta^{(2)}_{1}\big)^{-2}\int\mathscr{D}\bar{\Phi}_{\mu\nu}\mathscr{D}
         \Phi_{\mu\nu}\mathscr{D}\bar{\Phi}_{\mu}\mathscr{D}\Phi_{\mu}\mathscr{D}\bar{\Phi}\mathscr{D}
         \Phi\mathscr{D}\bar{\Lambda}_{\mu}\mathscr{D}\Lambda_{\mu}\mathscr{D}\bar{\Lambda}\mathscr{D}\Lambda\mathscr{D}\bar{c}\mathscr{D}c\times\\
         \times \delta[\Phi_{\mu}+\gamma_{\mu}c]\;\delta[\bar{\Phi}_{\mu}+\bar{c}\gamma_{\mu}]\;\delta[\Lambda]\;
         \delta[\bar{\Lambda}]\;\exp\left\{iS[\Phi_{\mu\nu},\bar{\Phi}_{\mu\nu},\Phi_{\mu},\bar{\Phi}_{\mu},\Phi,\bar{\Phi},\Lambda_{\mu},\bar{\Lambda}_{\mu},\Lambda,\bar{\Lambda}]\right\}
    \end{gathered}
\end{align}
and integrate over the field components annihilated by delta functions. This leads to
\begin{align}
    \begin{gathered}
         Z^{(2)}_{m}=\big(\Delta^{(2)}_{1}\big)^{-2}\int\mathscr{D}\bar{\Phi}_{\mu\nu}\mathscr{D}
         \Phi_{\mu\nu}\mathscr{D}\bar{\Phi}\mathscr{D}\Phi\mathscr{D}\bar{\Lambda}_{\mu}\mathscr{D}\Lambda_{\mu}\times\\
         \times \exp\left\{iS[\Phi_{\mu\nu},\bar{\Phi}_{\mu\nu},\Phi_{\mu}=0,\bar{\Phi}_{\mu}=0,
         \Phi,\bar{\Phi},\Lambda_{\mu},\bar{\Lambda}_{\mu},\Lambda=0,\bar{\Lambda}=0]\right\}.
    \end{gathered}
\end{align}
Here $S$ is the action \eqref{eq:76} expressed in terms of gamma-irreducible components of fields $\phi_{\mu\nu}$, $\lambda_{\mu}$,
\begin{align}
    \begin{gathered}
         S[\Phi_{\mu\nu},\bar{\Phi}_{\mu\nu},\Phi_{\mu}=0,\bar{\Phi}_{\mu}=0,\Phi,\bar{\Phi},\Lambda_{\mu},\bar{\Lambda}_{\mu},\Lambda=0,\bar{\Lambda}=0]=\\
         =i\int d^4x\sqrt{|g|}\left\{-2\bar{\Psi}^{\mu\nu}\left(\slashed{\nabla}_{2}+im\right)\Phi_{\mu\nu}-\frac{(d-1)!}{(d-5)!}
         \bar{\Psi}\left(\slashed{\nabla}_{0}+im\frac{d}{d-4}\right)\Phi\right\}-\\
         -ip^{(2)}\int d^4x\sqrt{|g|}\;\bar{\Lambda}^{\mu}\left(\slashed{\nabla}_{1}+im\frac{d-2}{d-4}\right)\Lambda_{\mu}.
    \end{gathered}
\end{align}

One can observe that the action becomes completely diagonal in gamma-irreducible components of different order. Finally, expressing remaining integrals in terms of determinants of the corresponding operators, we arrive at the final expression for the one-loop contribution to the functional integral of the model of antisymmetric rank-2 tensor spinor fields:
\begin{equation}
\label{Zr2}
Z^{(2)}_{m}=\frac{\Delta_{2}^{(2)}\Delta_{0}^{(2)}\Delta_{1}^{(2)}}{\big(\Delta_{1}^{(2)}\big)^2}=\frac{\Delta_{2}^{(2)}\Delta_{0}^{(2)}}{\Delta_{1}^{(2)}},\qquad \Delta_{p}^{(l)}=\text{Det}\left(i\slashed{\nabla}_{p}-m\frac{d-2p}{d-2l}\right).
\end{equation}
Operators $\slashed{\nabla}_{p}$ are defined as before, $\slashed{\nabla}_{p}=\gamma^{\mu}\nabla_{\mu}\pm i(1/2)\sqrt{r}(d-2p)$, and their determinants are evaluated on gamma-irreducible components of the corresponding rank $p$.

Taking the limit $m=0$ in the expression \eqref{Zr2}, one finds that $Z^{(2)}_{m=0}=\Delta_{2}\Delta_{0}\Delta_{1}^{-1}$, where $\Delta_{p}=\text{Det}(i\slashed{\nabla}_{p})$. It is easy to see that the relation $Z^{(2)}_{m=0}=Z^{(2)}Z^{(1)}$ holds, where $Z^{(p)}$ is a one-loop contribution of massless rank-p theory, obtained in \cite{BBKN2}. Thus, in the massless limit, the effective action (the logarithm of the partition function $Z$) of the originally massive rank-2 theory with Stueckelberg fields decomposes into the sum of effective actions of massless rank-2 and massless rank-1 theories and does not coincide with effective action of the originally massless theory ($\log Z^{(2)}$). The difference is due to the fact that introduction of Stueckelberg fields adds new degrees of freedom affecting the dynamics of the model. The contribution of Stueckelberg fields can be explicitly identified with $Z^{(1)}$ since the corresponding part of the action \eqref{eq:76} coincides with the action of rank-1 antisymmetric tensor spinor field.

\subsection{Quantization of the second-stage reducible theory with mass deformation}
We now turn to the example of the second-stage reducible gauge theory. The model of rank-3 massive totally antisymmetric tensor spinor field in $AdS_{d}$ space is described by the action
\begin{align}
    \begin{gathered}
         S[\psi_{\mu\nu\rho},\bar{\psi}_{\mu\nu\rho}]=i\int d^{d}x\;\sqrt{|g|}\;\bar{\psi}_{\mu_{1}\mu_{2}\mu_{3}}\gamma^{\mu_{1}\mu_{2}\mu_{3}\sigma\nu_{1}\nu_{2}\nu_{3}}
         D_{\sigma}\psi_{\nu_{1}\nu_{2}\nu_{3}}-\;\;\;\;\;\;\;\;\;\;\;\;\;\;\;\\\;\;\;\;\;\;\;\;\;\;\;\;\;\;\;\;\;\;\;\;\;\;\;\;\;
         \;\;\;\;\;\;\;\;\;\;\;\;\;\;\;\;\;\;\;\;\;\;\;\;\;
         -m\int d^{d}x\;\sqrt{|g|}\;\psi_{\mu_{1}\mu_{2}\mu_{3}}\gamma^{\mu_{1}\mu_{2}\mu_{3}\nu_{1}\nu_{2}\nu_{3}}\psi_{\nu_{1}\nu_{2}\nu_{3}},
    \end{gathered}
\end{align}
whose kinetic term is invariant under the gauge transformation $\delta\psi_{\mu\mu\mu}=D_{[\mu}f_{\nu\rho]}$. Here the same notation as in subsection 5.1 is used, but now spacetime dimension satisfies inequality $d>6$. As in the previous case, in order to restore gauge invariance of the mass term, we introduce Stueckelberg fields:
\begin{align}
    \begin{gathered}
         S[\psi,\bar{\psi},\lambda,\bar{\lambda}]=i\int d^{d}x\;\sqrt{|g|}\;\bar{\psi}_{\mu_{1}\mu_{2}\mu_{3}}\gamma^{\mu_{1}\mu_{2}\mu_{3}\sigma\nu_{1}\nu_{2}\nu_{3}}
         D_{\sigma}\psi_{\nu_{1}\nu_{2}\nu_{3}}-\\-m\int d^{d}x\;\sqrt{|g|}\;\left(\bar{\psi}_{\mu_{1}\mu_{2}\mu_{3}}-\frac{1}{m}\bar{\lambda}_{[\mu_{1}\mu_{2}}
         \overleftarrow{D}_{\mu_{3}]}\right)\gamma^{\mu_{1}\mu_{2}\mu_{3}\nu_{1}\nu_{2}\nu_{3}}
         \left(\psi_{\nu_{1}\nu_{2}\nu_{3}}-\frac{1}{m}D_{[\nu_{1}}\lambda_{\nu_{2}\nu_{3}]}\right).
    \end{gathered}
\end{align}
The resulting action is invariant under the transformation,
\begin{align}
    \begin{gathered}
         \delta\psi_{\mu\nu\rho}=D_{[\mu}f_{\nu\rho]},\qquad\delta\lambda_{\mu\nu}=mf_{\mu\nu}+D_{[\mu}g_{\nu]},\label{eq:109}
    \end{gathered}
\end{align}
and can be brought to the form
\begin{align}
    \begin{gathered}
         S[\psi,\bar{\psi},\lambda,\bar{\lambda}]
         =S_{0}\mp i\frac{\sqrt{r}}{m}(d-5)S_{St}+S_{ND},
    \end{gathered}
\end{align}
where
\begin{itemize}
\item{$S_{0}$ denotes the action of the model prior to introduction of Stueckelberg fields:}
$$S_{0}=i\int d^{d}x\;\sqrt{|g|}\;\bar{\psi}_{\mu_{1}\mu_{2}\mu_{3}}\gamma^{\mu_{1}\mu_{2}\mu_{3}\sigma\nu_{1}\nu_{2}\nu_{3}}D_{\sigma}\psi_{\nu_{1}\nu_{2}\nu_{3}}-m\int d^{d}x\;\sqrt{|g|}\;\bar{\psi}_{\mu_{1}\mu_{2}\mu_{3}}\gamma^{\mu_{1}\mu_{2}\mu_{3}\nu_{1}\nu_{2}\nu_{3}}\psi_{\nu_{1}\nu_{2}\nu_{3}}$$
\item{$S_{St}$ is the diagonal contribution of Stueckelberg fields:}
$$S_{St}=\int d^{d}x\;\sqrt{|g|}\;\bar{\lambda}_{\mu_{1}\mu_{2}}\gamma^{\mu_{1}\mu_{2}\nu_{1}\nu_{2}\nu_{3}}D_{\alpha}\lambda_{\beta\gamma}$$
\item{$S_{ND}$ represents non-diagonal contributions:}
$$
S_{ND}=\int d^{d}x\;\sqrt{|g|}\;\bar{\lambda}_{\mu_1\mu_2}\overleftarrow{D}_{\mu_3}\gamma^{\mu_{1}\mu_{2}\mu_{3}\nu_{1}\nu_{2}\nu_{3}}\psi_{\nu_{1}\nu_{2}\nu_{3}}+\int d^{d}x\;\sqrt{|g|}\;\bar{\psi}_{\mu_{1}\mu_{2}\mu_{3}}\gamma^{\mu_{1}\mu_{2}\mu_{3}\nu_{1}\nu_{2}\nu_{3}}D_{\nu_1}\lambda_{\nu_2\nu_3}
$$
\end{itemize}
In order to eliminate derivative cross terms, we perform the field redefinition
\begin{align}
    \begin{gathered}
         \psi_{\mu\nu\rho}=\phi_{\mu\nu\rho}+\frac{i}{d-6}\gamma_{[\mu}\lambda_{\nu\rho]}
    \end{gathered}
\end{align}
after which the action takes the form
\begin{align}
    \begin{gathered}
         S[\phi,\bar{\phi},\lambda,\bar{\lambda}]=\tilde{S}_{0}-\frac{p^{(3)}}{m}\tilde{S}_{St}+p^{(3)}\tilde{S}_{ND},\qquad p^{(3)}\equiv(d-5)\left(\pm\sqrt{r}+\frac{m}{d-6}\right),
    \end{gathered}
\end{align}
where
\begin{itemize}
\item{$\tilde{S}_{0}$ is the diagonal contribution of the field $\phi$:}
$$\tilde{S}_{0}=i\int d^{d}x\;\sqrt{|g|}\;\bar{\phi}_{\mu_{1}\mu_{2}\mu_{3}}\gamma^{\mu_{1}\mu_{2}\mu_{3}\sigma\nu_{1}\nu_{2}\nu_{3}}D_{\sigma}\phi_{\nu_{1}\nu_{2}\nu_{3}}-m\int d^{d}x\;\sqrt{|g|}\;\bar{\phi}_{\mu_{1}\mu_{2}\mu_{3}}\gamma^{\mu_{1}\mu_{2}\mu_{3}\nu_{1}\nu_{2}\nu_{3}}\phi_{\nu_{1}\nu_{2}\nu_{3}}$$
\item{$\tilde{S}_{St}$ is the diagonal contribution of Stueckelberg fields including the additional term arising after substitution:}
$$\tilde{S}_{St}=i\int d^{d}x\;\sqrt{|g|}\;\bar{\lambda}_{\mu_{1}\mu_{2}}\gamma^{\mu_{1}\mu_{2}\nu_{1}\nu_{2}\nu_{3}}D_{\nu_{1}}\lambda_{\nu_{2}\nu_{3}}+m\frac{d-4}{d-6}\int d^{d}x\;\sqrt{|g|}\;\bar{\lambda}_{\mu_{1}\mu_{2}}\gamma^{\mu_{1}\mu_{2}\nu_{1}\nu_{2}}\lambda_{\mu_{1}\mu_{2}}$$
\item{$\tilde{S}_{ND}$ denotes non-diagonal contributions after substitution:}
$$\tilde{S}_{ND}=i\int d^{d}x\;\sqrt{|g|}\;\bar{\lambda}_{\mu_{1}\mu_{2}}\gamma^{\mu_{1}\mu_{2}\nu_{1}\nu_{2}\nu_{3}}\phi_{\nu_{1}\nu_{2}\nu_{3}}-i\int d^{d}x\;\sqrt{|g|}\;\bar{\phi}_{\mu_{1}\mu_{2}\mu_{3}}\gamma^{\mu_{1}\mu_{2}\mu_{3}\nu_{1}\nu_{2}}\lambda_{\nu_{1}\nu_{2}}.$$
\end{itemize}

Non-derivative cross terms can be expressed in terms of gamma-irreducible components of fields as follows:
\begin{align}
    \begin{gathered}
         \bar{\lambda}_{\nu\rho}\gamma^{\nu\rho\alpha\beta\gamma}\phi_{\alpha\beta\gamma}=2\frac{(d-4)!}{(d-5)!}\bar{\Lambda}_{\mu\nu}\Phi^{\mu\nu}+\frac{(d-2)!}{(d-5)!}\bar{\Lambda}_{\mu}\Phi^{\mu}+\frac{d!}{(d-5)!}\bar{\Lambda}\Phi,
    \end{gathered}
\end{align}
where
\begin{align}
    \begin{gathered}
         \phi_{\mu\nu\rho}=\Phi_{\mu\nu\rho}+\gamma_{[\mu}\Phi_{\nu\rho]}+\gamma_{[\mu\nu}\Phi_{\rho]}+\gamma_{\mu\nu\rho}\Phi,\;\;\;\;\;\;\;\;\gamma^{\mu}\Phi_{\mu\nu\rho}=0,\;\gamma^{\mu}\Phi_{\mu\nu}=0,\;\gamma^{\mu}\Phi_{\mu}=0\\
         \lambda_{\mu\nu}=\Lambda_{\mu\nu}+\gamma_{[\mu}\Lambda_{\nu]}+\gamma_{\mu\nu}\Lambda,\;\;\;\;\;\;\;\;\gamma^{\mu}\Lambda_{\mu\nu}=0,\;\gamma^{\mu}\Lambda_{\mu}=0.
    \end{gathered}
\end{align}
Therefore, similarly to the previous case, one has to choose gauge conditions that eliminate cross terms. Consider the operators
\begin{align}
    \begin{gathered}
         {P_{2,0}}_{\mu\nu}(\phi_{\alpha\beta\gamma})=\gamma^{\rho}\phi_{\mu\nu\rho}+\frac{1}{d-2}\gamma_{[\mu}\gamma^{\alpha\beta}\phi_{\nu]\alpha\beta},
    \end{gathered}\\
    \begin{gathered}
         {P_{1}}_{\mu}(\lambda_{\alpha\beta})=\gamma^{\nu}\lambda_{\mu\nu}-\frac{1}{d}\gamma_{\mu}\gamma^{\alpha\beta}\lambda_{\alpha\beta}.\;\;\;\;\;\;\;
    \end{gathered}
\end{align}
In terms of gamma-irreducible components, these operators act as follows:
\begin{align}
    \begin{gathered}
         {P_{2,0}}_{\mu\nu}(\phi_{\alpha\beta\gamma})=\frac{1}{3}(d-4)\Phi_{\mu\nu}-\gamma_{\mu\nu}\Phi,
    \end{gathered}\\
    \begin{gathered}
         {P_{1}}_{\mu}(\lambda_{\alpha\beta})=\frac{1}{2}(d-2)\Lambda_{\mu}.\;\;\;\;\;\;\;\;
    \end{gathered}
\end{align}
Thus, if one imposes gauge conditions
\begin{align}
    \begin{gathered}
         {P_{2,0}}_{\mu\nu}(\phi_{\alpha\beta\gamma})={P_{2,0}}_{\mu\nu}\left(\psi_{\alpha\beta\gamma}-\frac{i}{d-6}\gamma_{[\alpha}\lambda_{\beta\gamma]}\right)=0,\qquad
         {P_{1}}_{\mu}(\lambda_{\alpha\beta})=0,
    \end{gathered}
\end{align}
which project out from the field $\phi$ its components of the zeroth and second order, and from the field $\lambda$ the component of the first order, then after integration with the corresponding delta functions, the non-derivative cross terms vanish.

In contrast to the previous case, the model under consideration is a second-stage reducible gauge theory, since the symmetry \eqref{eq:109} is now invariant under the transformation
\begin{align}
    \begin{gathered}
         \delta f_{\mu\nu}=D_{[\mu}{f_{1}}_{\nu]},\qquad \delta g_{\mu}=-m{f_{1}}_{\mu}+D_{\mu}g_{1},\label{eq:120}
    \end{gathered}
\end{align}
which, in its turn, is invariant under
\begin{align}
    \begin{gathered}
         \delta{f_{1}}_{\mu}=D_{\mu}f_{2},\qquad \delta g_{1}=m{f_{2}}.
    \end{gathered}
\end{align}
At the same time, the operators introduced above satisfy the properties
\begin{align}
    \begin{gathered}
         {P_{1}}_{\mu}({P_{2,0}}_{\alpha\beta})=0,\;\;\;\;\;\gamma^{\mu}{P_{1}}_{\mu}=0.
    \end{gathered}
\end{align}

Following the general scheme, we insert into the integral
\begin{align}
    \begin{gathered}
         Z=\int\mathscr{D}\bar{\psi}\mathscr{D}\psi\mathscr{D}\bar{\lambda}\mathscr{D}\lambda\;e^{iS[\bar{\psi},\psi,\bar{s},s]}
    \end{gathered}
\end{align}
the functional unity
\begin{align}
    \begin{gathered}
         1=\Delta^{-1}\int \mathscr{D}\mu_{f,g}\;\hat{\hat{\delta}}\left[{P_{2,0}}_{\mu\nu}\left(\phi_{\alpha\beta\gamma}
         +D_{[\alpha}f_{\beta\gamma]}-i\frac{m}{d-6}\gamma_{[\alpha}f_{\beta\gamma]}-i\frac{1}{d-6}
         \gamma_{[\alpha}D_{\beta}g_{\gamma]}\right)\right]\times\;\;\\
         \;\;\;\;\;\;\;\;\;\;\;\;\;\;\;\;\;\;\;\;\;\;\;\;\;\;\times\hat{\delta}\left[{P_{1}}_{\mu}
         \left(\lambda_{\alpha\beta}+mf_{\alpha\beta}+D_{[\alpha}g_{\beta]}\right)\right],\label{eq:124}
    \end{gathered}
\end{align}
and an analogous unity for conjugated fields.

We now define the modified integration measure. We fix the symmetry \eqref{eq:120} by conditions
\begin{align}
    \begin{gathered}
         {P_{1}}_{\mu}\left(f_{\alpha\beta}+i\frac{1}{d-6}\gamma_{[\alpha}g_{\beta]}\right)=0,\qquad\gamma^{\mu}g_{\mu}=0
    \end{gathered}
\end{align}
and insert into the naively defined integral
\begin{align}
    \begin{gathered}
         \int \mathscr{D}f\mathscr{D}g\;\delta\left[{P_{2,0}}_{\mu\nu}\left(\phi_{\alpha\beta\gamma}+D_{[\alpha}f_{\beta\gamma]}-i\frac{m}{d-6}\gamma_{[\alpha}f_{\beta\gamma]}-i\frac{1}{d-6}\gamma_{[\alpha}D_{\beta}g_{\gamma]}\right)\right]\times\;\;\\
         \;\;\;\;\;\;\;\;\;\;\;\;\;\;\;\;\;\;\;\;\;\;\;\;\;\;\times\delta\left[{P_{1}}_{\mu}\left(\lambda_{\alpha\beta}+mf_{\alpha\beta}+D_{[\alpha}g_{\beta]}\right)\right]\label{eq:126}
    \end{gathered}
\end{align}
the functional unity
\begin{align}
    \begin{gathered}
         1=\Delta_{m}^{-1}\int \mathscr{D}\mu_{f_{1},g_{1}}\;\hat{\delta}\left[{P_{1}}_{\mu}\left(f_{\alpha\beta}
         +i\frac{1}{d-6}\gamma_{[\alpha}g_{\beta]}+D_{[\alpha}{f_{1}}_{\beta]}-i\frac{m}{d-6}
         \gamma_{[\alpha}{f_{1}}_{\beta]}+i\frac{1}{d-6}\gamma_{[\alpha}D_{\beta]}g_{1}\right)\right]\times\;\;\;\;\;\\
         \;\;\;\;\;\;\;\;\;\;\;\;\;\;\;\;\;\;\;\;\;\;\;\;\;\;\times\delta\left[\gamma^{\mu}\left(g_{\mu}-m{f_{1}}_{\mu}
         +D_{\mu}g_{1}\right)\right]. \label{eq:127}
    \end{gathered}
\end{align}

It is straightforward to see that this expression is completely analogous to \eqref{eq:85}. Therefore, proceeding in the same way as in the previous case, we obtain
\begin{align}
    \begin{gathered}
        \Delta_{m}=\text{Det}\left(i\slashed{\nabla}_{1}+m\frac{d-2}{d-6}\right)\equiv\tilde{\Delta}^{(3)}_{1},\qquad
        \tilde{\Delta}^{(l)}_{p}=\text{Det}\left(i\slashed{\nabla}_{p}+m\frac{d-2p}{d-2l}\right),
    \end{gathered}\\
    \begin{gathered}
         \hat{\delta}[{P_{1}}_{\mu}(F_{\alpha\beta})]=\int\mathscr{D}\bar{\pi}\;e^{i\bar{\pi}^{\mu}{P_{1}}_{\mu}(F_{\alpha\beta})}\;
         \delta[\bar{\pi}^{\mu}\gamma_{\mu}]=\int\mathscr{D}c_{1}\;\delta[{P_{1}}_{\mu}(F_{\alpha\beta})+\gamma_{\mu}c_{1}],
    \end{gathered}
\end{align}
where $F_{\alpha\beta}=f_{\alpha\beta}-i\frac{1}{d-6}\gamma_{[\alpha}g_{\beta]}$. Now, substituting the fully defined unity \eqref{eq:127} into the integral \eqref{eq:126} and making the transformations analogous to \eqref{eq:40}, we obtain the expression for the modified integration measure
\begin{align}
    \begin{gathered}
         \mathscr{D}\mu_{f,g}=\mathscr{D}f\mathscr{D}g\;\big(\tilde{\Delta}^{(3)}_{1}\big)^{-1}\int \mathscr{D}c_{1}\delta\left[{P_{1}}_{\mu}\left(f_{\alpha\beta}+i\frac{1}{d-6}
         \gamma_{[\alpha}g_{\beta]}\right)+\gamma_{\mu}c_{1}\right]\delta[\gamma^{\mu}g_{\mu}]. \label{eq:130}
    \end{gathered}
\end{align}

Modified delta functions in the integral \eqref{eq:124} are defined in complete analogy with a general case, since no mixing of integration parameters occurs. In addition, no determinants of differential operators arise in the computation, so that all such factors can be omitted. Exponential in the expression for the delta function of auxiliary fields
\begin{align}
    \begin{gathered}
         \delta[{P_{1}}_{\mu}(\lambda_{\alpha\beta})]=\int \mathscr{D}\bar{\eta}\;e^{i\bar{\eta}^{\mu}{P_{1}}_{\mu}(\lambda_{\alpha\beta})}
    \end{gathered}
\end{align}
is invariant under the transformation $\delta\bar{\eta}_{\mu}=\bar{\chi}\gamma_{\mu}$. We fix this symmetry by the gauge condition $\bar{\eta}^{\mu}\gamma_{\mu}=0$. Repeating the steps analogous to those described above, we obtain
\begin{align}
    \begin{gathered}
         \hat{\delta}[{P_{1}}_{\mu}(\lambda_{\alpha\beta})]=\int\mathscr{D}c_{2}\;\delta[{P_{1}}_{\mu}(\lambda_{\alpha\beta})+\gamma_{\mu}c_{2}]
         =\int\mathscr{D}\bar{\pi}\;e^{i\bar{\pi}^{\mu}{P_{1}}_{\mu}(\lambda_{\alpha\beta})}\delta[\bar{\pi}^{\nu}\gamma_{\nu}].\label{eq:133}
    \end{gathered}
\end{align}
Exponential in the expression for the delta function of the field $\psi_{\mu\nu\rho}$
\begin{align}
    \begin{gathered}
         \delta[{P_{2,0}}_{\mu\nu}(\phi_{\alpha\beta\gamma})]=\int\mathscr{D}\bar{\pi}\;e^{i\bar{\pi}^{\mu\nu}{P_{2,0}}_{\mu\nu}(\phi_{\alpha\beta\gamma})}
    \end{gathered}
\end{align}
is invariant under the transformation
\begin{align}
    \begin{gathered}
         \delta\bar{\pi}_{\mu\nu}=\bar{\xi}_{\alpha}{P_{1}}^{\alpha}_{\mu\nu}\equiv\bar{\xi}_{\alpha}
         \left(\gamma_{[\nu}\delta^{\alpha}_{\mu]}+\frac{1}{d}\gamma^{\alpha}\gamma_{\mu\nu}\right),\label{eq:135}
    \end{gathered}
\end{align}
which in its turn is invariant under $\delta\bar{\xi}_{\mu}=\bar{\xi}_{1}\gamma_{\mu}$. We fix the symmetry \eqref{eq:135} by the gauge condition
\begin{align}
    \begin{gathered}
         (\bar{\pi}_{\alpha\beta})\overleftarrow{{P_{1}}}_{\mu}\equiv\overline{{P_{1}}_{\mu}(\pi_{\alpha\beta})}=0
    \end{gathered}
\end{align}
and the symmetry $\delta\bar{\xi}_{\mu}=\bar{\xi}_{1}\gamma_{\mu}$ by the gauge condition $\bar{\xi}_{\mu}\gamma^{\mu}=0$.
In doing so, when fixing the symmetry \eqref{eq:135}, one should take into account that the exponential in the expression for corresponding delta function
\begin{align}
    \begin{gathered}
         \delta[(\bar{\pi}_{\alpha\beta})\overleftarrow{{P_{1}}}_{\mu}]=\int \mathscr{D}\theta\;e^{i(\bar{\pi}_{\alpha\beta})\overleftarrow{{P_{1}}}_{\mu}\theta^{\mu}}
    \end{gathered}
\end{align}
is invariant under $\delta\theta_{\mu}=\gamma_{\mu}\theta_{1}$. This symmetry is fixed by the gauge condition $\gamma^{\mu}\theta_{\mu}=0$. Further, repeating the steps analogous to \eqref{eq:54} (omitting determinants of ultralocal operators), we have
\begin{align}
    \begin{gathered}
         \hat{\hat{\delta}}[{P_{2,0}}_{\mu\nu}(\phi_{\alpha\beta\gamma})]=\int\mathscr{D}c\;\delta[{P_{2,0}}_{\mu\nu}(\phi_{\alpha\beta\gamma})
         +\gamma_{[\mu}c_{\nu]}]\;\delta[\gamma^{\mu}c_{\mu}]=\\
         =\int\mathscr{D}\bar{\pi}e^{i\bar{\pi}^{\mu\nu}{P_{2,0}}_{\mu\nu}(\phi_{\alpha\beta\gamma})}
         \int\mathscr{D}\bar{c}'\delta[(\bar{\pi}_{\alpha\beta})\overleftarrow{{P_{1}}}_{\rho}+\bar{c}'\gamma_{\rho}]. \label{eq:140}
    \end{gathered}
\end{align}
We now substitute expressions \eqref{eq:130}, \eqref{eq:133} and \eqref{eq:140} into the integral \eqref{eq:124}:
\begin{align}
    \begin{gathered}
         \Delta=\big(\tilde{\Delta}^{(3)}_{1}\big)^{-1}\int\mathscr{D}f\mathscr{D}g\int\mathscr{D}c_{1}
         \delta\left[{P_{1}}_{\mu}\left(f_{\alpha\beta}+i\frac{1}{d-6}\gamma_{[\alpha}g_{\beta]}\right)
         +\gamma_{\mu}c_{1}\right]\delta[\gamma^{\mu}g_{\mu}]\times\\
         \times\int\mathscr{D}c\;\delta\left[{P_{2,0}}_{\mu\nu}\left(\phi_{\alpha\beta\gamma}+D_{[\alpha}f_{\beta\gamma]}
         -i\frac{m}{d-6}\gamma_{[\alpha}f_{\beta\gamma]}-i\frac{1}{d-6}\gamma_{[\alpha}D_{\beta}g_{\gamma]}\right)
         +\gamma_{[\mu}c_{\nu]}\right]\;\delta[\gamma^{\mu}c_{\mu}]\times\\
         \times\int\mathscr{D}c_{2}\;\delta\left[{P_{1}}_{\mu}\left(\lambda_{\alpha\beta}+m\left(f_{\alpha\beta}
         +i\frac{1}{d-6}\gamma_{[\alpha}g_{\beta]}\right)+D_{[\alpha}g_{\beta]}-i\frac{m}{d-6}\gamma_{[\alpha}g_{\beta]}\right)+\gamma_{\mu}c_{2}\right].
    \end{gathered}
\end{align}
The delta function $\delta[\gamma^{\mu}g_{\mu}]$ eliminates zeroth-order gamma-irreducible component of $g_{\mu}$, while the operator $P_{2,0}$ annihilates first-order component of the field it acts upon. This implies that under the action of $P_{2,0}$ one can freely add terms of the form $\gamma_{[\alpha\beta}g_{\gamma]},\;\gamma^{\alpha}g_{\alpha}=0$, and thus one can again make the substitution
\begin{align}
    \begin{gathered}
         F_{\alpha\beta}=f_{\alpha\beta}+i\frac{1}{d-6}\gamma_{[\alpha}g_{\beta]},
    \end{gathered}
\end{align}
which removes mixing the integration variables. Next, as in the previous case, we represent delta functions as integrals and decompose integration variables into gamma-irreducible components. Only diagonal contributions of second and zeroth order from the operator $P_{2,0}$ and first-order contributions from the operator $P_{1}$ survive. As a result, one obtains
\begin{align}
    \begin{gathered}
         \Delta\;=\frac{\tilde{\Delta}^{(3)}_{2}\tilde{\Delta}^{(3)}_{0}\tilde{\Delta}^{(3)}_{1}}{\tilde{\Delta}^{(3)}_{1}}
         =\tilde{\Delta}^{(3)}_{2}\tilde{\Delta}^{(3)}_{0}.
    \end{gathered}
\end{align}

Now, in complete analogy with what was above, we substitute obtained expressions into the functional integral and separate integration over first-stage gauge group
\begin{align}
    \begin{gathered}
         Z^{(3)}_{m}=\frac{1}{\Delta^{2}}\int\mathscr{D}\mu
         \mathscr{D}\bar{\mu}\;\delta[(d-4)\Phi_{\mu\nu}-3\gamma_{\mu\nu}\Phi+\gamma_{[\mu}C_{\nu]}+\gamma_{\mu\nu}C]\;\delta[(d-4)\bar{\Phi}_{\mu\nu}
         +3\bar{\Phi}\gamma_{\mu\nu}+\bar{C}_{[\mu}\gamma_{\nu]}-\bar{C}\gamma_{\mu\nu}]\times\\
         \times\delta[C]\;\delta[\bar{C}]\;\delta[\Lambda_{\mu}+\gamma_{\mu}c_{2}]\;\delta[\bar{\Lambda}_{\mu}+\bar{c}_{2}\gamma_{\mu}]\;\exp[iS]=\\=
         \frac{1}{\Delta^{2}}\int\mathscr{D}\Phi_{\mu\nu\rho}\mathscr{D}\Phi_{\mu}\mathscr{D}\Lambda_{\mu\nu}\mathscr{D}\Lambda     \mathscr{D}\bar{\Phi}_{\mu\nu\rho}\mathscr{D}\bar{\Phi}_{\mu}\mathscr{D}\bar{\Lambda}_{\mu\nu}\mathscr{D}\bar{\Lambda}\;\exp[i\tilde{S}],
    \end{gathered}
\end{align}
where
\begin{itemize}
\item{$\mathscr{D}\mu\equiv\mathscr{D}\Phi_{\mu\nu\rho}\mathscr{D}\Phi_{\mu\nu}\mathscr{D}\Phi_{\mu}
\mathscr{D}\Phi\mathscr{D}\Lambda_{\mu\nu}\mathscr{D}\Lambda_{\mu}\mathscr{D}\Lambda\mathscr{D}C_{\mu}\mathscr{D}C\mathscr{D}c_{2}$},
\item{$\mathscr{D}\bar{\mu}=\mathscr{D}\bar{\Phi}_{\mu\nu\rho}\mathscr{D}\bar{\Phi}_{\mu\nu}
\mathscr{D}\bar{\Phi}_{\mu}\mathscr{D}\bar{\Phi}\mathscr{D}\bar{\Lambda}_{\mu\nu}
\mathscr{D}\bar{\Lambda}_{\mu}\mathscr{D}\bar{\Lambda}\mathscr{D}\bar{C}_{\mu}\mathscr{D}\bar{C}\mathscr{D}\bar{c}_{2}$},
\item{$\tilde{S}=S[\Phi_{\mu\nu\rho},\Phi_{\mu\nu}=0,\Phi_{\mu},\Phi=0,\Lambda_{\mu\nu},\Lambda_{\mu}=0,
\Lambda,\bar{\Phi}_{\mu\nu\rho},\bar{\Phi}_{\mu\nu}=0,\bar{\Phi}_{\mu},\bar{\Phi}=0,\bar{\Lambda}_{\mu\nu},\bar{\Lambda}_{\mu}=0,\bar{\Lambda}].$}
\end{itemize}
Expressing Gaussian integrals in terms of determinants of the corresponding operators, one arrives at the final expression for the effective action of the model of antisymmetric rank-3 fermionic field:
\begin{align}
    \begin{gathered}
         Z^{(3)}_{m}=\frac{\tilde{\Delta}^{(3)}_{3}\tilde{\Delta}^{(3)}_{1}\tilde{\Delta}^{(3)}_{2}
         \tilde{\Delta}^{(3)}_{0}}{\big(\tilde{\Delta}^{(3)}_{2}\big)^2\big(\tilde{\Delta}^{(3)}_{0}\big)^2}
         =\frac{\tilde{\Delta}^{(3)}_{3}\tilde{\Delta}^{(3)}_{1}}{\tilde{\Delta}^{(3)}_{2}\tilde{\Delta}^{(3)}_{0}}.
    \end{gathered}
\end{align}

As in the previous case, in the massless limit the resulting effective action decomposes into a sum of massless effective actions corresponding to the original and Stueckelberg fields, $Z^{(3)}_{m=0}=Z^{(3)}Z^{(2)}$. One observes that in both cases the contribution of Stueckelberg fields corresponds to the effective action of a theory with the rank lower by one as compared to the rank of the original field.

In contrast to the general scheme, in this section non-diagonal gauge-fixing conditions were employed, which allowed us to explicitly separate contributions of the original and Stueckelberg fields and to obtain the final expression for the effective action.

\section{Conclusions}
 \label{Sect:conclusions}
We considered a general gauge theory
with linearly dependent generators and deformed it by mass or/and
interaction terms that violate its gauge invariance and studied
quantization of this theory. The peculiarity of such a theory is
that its kinetic term in the action is gauge invariant, but the terms
causing deformation are not. This circumstance may lead to certain
problems in constructing quantum effective action, in
particular, manifestly covariant Schwinger-DeWitt technique for its calculation
might turn out to be not directly applicable.

In case of Abelian non-deformed theory we developed generic
Stueckelberg-type procedure that allowed us to
convert the theory with broken symmetry into a gauge invariant theory subject to
quantization in terms of the original
fields and Stueckelberg gauge fields within the
procedure recently proposed in \cite{BBKN1}. As a result, we derived
the functional integrals for partition functions, equipped with the necessary
set of ghost fields. This was done for the class of theories whose original non-deformed
version belonged to the first and second stages of reducibility.

As an application of this general approach, we considered
quantization of a recently proposed massive fermionic totally
antisymmetric tensor field theory in $AdS$ space
(fermionic $p$-form gauge theory) \cite{BKR}. In this case, the massless part
of the action is gauge invariant, and the massive part is a
deformation. At arbitrary $p$, the original massless theory is the one
of $p-1$ stages of reducibility. We have analyzed two classes of the
deformed theory corresponding to the first and
second stages and derived the corresponding functional integral
representations for their effective actions. In both cases, all fields including
the Stueckelberg ones can be integrated out and the one-loop effective
actions can be explicitly reduced to the contributions of functional determinants
of special Dirac-type differential operators acting on $p$-forms in $AdS$ space.

It is interesting to compare our consideration with the approach of the work \cite{KT} where a large number of different concrete bosonic deformed gauge models were investigated also using the Stueckulberg trick. In that work, the gauge transformations were taken to be irreducible while in our paper they are considered as reducible. To clarify this point we briefly discuss the aspects of quantization of bosonic and fermionic massive p-forms in general.
The main difference is that bosonic theories can be described consistently in arbitrary curved spacetime, while fermions can only be described in AdS, in other respects they are similar. Both bosonic and fermionic massive p-forms are formulated as ther Stueckelberg gauge theories with p-1 stages of reducibilities
(see, e.g. \cite{Buchbinder:2008kw} for the bosonic case). Then we can partially fix the symmetry and reduce the symmetry reducibility by stopping at any intermediate stage. This will lead to formally different but obvious equivalent formulations of massive p-forms both in bosonic and in fermionic cases
After such partial gauge fixing we can quantize any of these formally different formulations, but it is clear that we obtain equivalent quantum theories for both these formulations.  In the paper \cite{KT} it was considered the theories where partial gauge fixing has been performed in such a way that the remaining gauge symmetry for massive bosonic p-forms turned out to be irreducible. In the present paper we have studied quantization of massive fermionic 2-form and 3-form with a full number of reducibility stages. The use of one or another of the equivalent formulations is a matter of convenience when applied. In our paper, when computing effective action of deformed fermionic p-forms, we find it convenient to retain reducibility of undeformed theory and choose reducible form of gauge transformation and reducible gauge-fixing conditions. The reason is that additional delta functions appearing in construction of modified delta functions and modified integration measures make it possible to express the Faddeev–Popov determinants in terms of operators acting in space of gamma-irreducible fermionic fields.

The results obtained here open up the possibilities for studying various
aspects of quantization of deformed gauge theories and the construction of
their effective action. In particular, it seems
interesting to explore the following issues:
\begin{itemize}
\item{Constructing the Stueckelberg-type trick for general non-Abelian deformed
gauge theories and
developing the quantization procedure for these cases}
\item{Deriving the effective action for arbitrary deformed fermionic
$p$-form gauge theory with
any stage of reducibility;}
\item{Carrying out canonical quantization of deformed fermionic
$p$-form gauge theory;}
\item{Calculating the functional determinants in the effective
action of these deformed fermionic $p$-form gauge theories;}
\item{Supesymmetrization of the deformed fermionic $p$-form gauge
theory and studying their effective action.}
\end{itemize}
We plan to explore these issues in forthcoming works.

\section*{Acknowledgments}
The authors are grateful to S.M. Kuzenko for useful comments. The A.A. Averianov and A. O. Barvinsky thank the
``BASIS'' Foundation for the Advancement of Theoretical Physics and Mathematics for support.



\begin {thebibliography}{99}

\bibitem{GSchW}
M.~B.~Green, J.~H.~Schwarz, E.~Witten, ``Superstring Theory'', vols. 1 and 2, Cambridge Univ. Press, 1987.

\bibitem{P}
J.~Polchinski, ``String Theory'', vols. 1 and 2, Cambridge Univ. Press, 1998.

\bibitem{J}
C.~V.~Johnson, ``D-Branes'', Cambridge Univ. Press, 2003, 548 p.

\bibitem{O}
T.~Ortin, ``Gravity and Strings'', Cambridge Univ. Press, 2004.

\bibitem{BBSch}
K.~Becker, M.~Becker, J.~Y.~Schwarz, ``String Theory and M-Theory: A Modern Introduction'', Cambridge Univ. Press, 2007.

\bibitem{FVP}
D.~Z.~Freedman, A.~Van Proeyen, ``Supergravity'', Cambridge Univ. Press, 2012.

\bibitem{BLT}
R.~Blumenhagen, D.~Luest, S.~Theisen, ``Basic Concepts of String Theory'', Springer, 2013.

\bibitem{Ta}
Y.~Tanii, ``Introduction to Supergravity'', Springer, 2014.

\bibitem{HL}
S.~E.~Hjelmeland, U.~Lindstrom, ``Duality for non-specialist'', {\tt arXiv:hep-th/9705122}.

\bibitem{KT}
S.~M.~Kuzenko, K.~Turner, ``Effective actions for dual massive (super) $p$-forms'', JHEP \textbf{01} (2021) 040, {\tt arXiv:2009.08263 [hep-th]},

\bibitem{H}
A.~Hell, ``On the duality of massive Kalb-Ramond and Proca fields'', JCAP \textbf{01} (2022) 056, {\tt arXiv:2109.050302 [hep-th]}.

\bibitem{HO}
A.~Hell, I.~Obata, ``On the Kalb-Ramond field with non-minimal coupling to gravity'', {\tt arXiv:2602.21675 [hep-th]}.

\bibitem{BKW}
A.~O.~Barvinsky, A.~E.~Kalugin, W.~Wachowski, ``Schwinger-DeWitt expansion for the heat kernel of nonminimal operators in causal theories'',
Phys.~Rev. D \textbf{112} (2025) 076032, {\tt arXiv:2508.06439 [hep-th]}.


\bibitem{RR}
H.~Ruegg, M.~Ruiz-Altaba, ``The Stueckelberg field'', Int.~J.~Mod.~Phys. A \textbf{19} (2004) 3265, {\tt arXiv:hep-th/0304245}.

\bibitem{BBS}
I.~L.~Buchbinder, G.~de Berredo-Peixoto, I.~L.~Shapiro, ``Quantum effects in softly broken gauge theories in curved spacetimes'', Phys.~Lett. B \textbf{649} (2007) 454, {\tt arXiv:hep-th/0703189}.

\bibitem{BKP}
I.~L.~Buchbinder, E.~N.~Kirillova, N.~G.~Pletnev, ``Quantum equivalence of massive antisymmetric tensor field models in curved space'', Phys.~Rev. D \textbf{78} (2008) 084024, {\tt arXiv:0806.3505 [hep-th]}.

\bibitem{BVOS}
I.~L.~Buchbinder, P.~R.~B.~R.~do Vale, G.~Y.~Oyadomari, I.~L.~Shapiro,
``On the renormalization of massive vector field theory coupled to scalar in curved space-time'', Phys.~Rev., D \textbf{110} (2024) 12505, {\tt arXiv:2410.00991 [hep-th]}.

\bibitem{BKR}
I.~L.~Buchbinder, V.~A.~Krykhtin, L.~L.~Ryskina, ``Lagrangian formulation of massive fermionic totally antisymmetric tensor field theory in AdS(d) space'',
Nucl.~Phys. B \textbf{819} (2009), 453-477, {\tt arXiv:0902.1471 [hep-th]}.

\bibitem{Z}
Yu.~M.~Zinoviev, ``Note on antisymmetric spin-tensors'',
JHEP \textbf{04} (2009), 035, {\tt arXiv:0903.0262 [hep-th]}.

\bibitem{CFMS}
A.~Campoleoni, D.~Francia, J.~Mourad, A.~Sagnotti, ``Unconstrained
higher spins of mixed symmetry. II. Fermi fields'', Nucl.~Phys. B \textbf{828} (2010) 405, 
{\tt arXiv:0904.4447 [hep-th]}.

\bibitem{H-1}
C.~M.~Hull, ``Strongly coupled gravity and dualities'', Nucl.~Phys. B
\textbf{583} (2000) 237, {\tt arXiv:hep-th/0004195}. 

\bibitem{H-2}
C.~M.~Hull, ``Symmetries and compactifications of (4,0) conformal
gravity'', JHEP \textbf{12} (2000) 007, {\tt arXiv:hep-th/0011215}.

\bibitem{H-3}
C.~M.~Hull, ``Duality in gravity and higher spin gauge fields'',
JHEP \textbf{09} (2001) 027, {\tt arXiv:hep-th/0107149}.

\bibitem{W}
P.~West, ``$E(11)$ and M theory'', Class. Quant. Grav. \textbf{18}
(2001) 443, {\tt arXiv:hep-th/0104081}.

\bibitem{B}
L.~Borsten, ``$D=6,\, {\cal N}=(2,0)$ and ${\cal N}=(4,0)$
theories'', Phys.~Rev. D \textbf{97} (2018) 066014, {\tt
arXiv:1708.02573 [hep-th]}.

\bibitem{HLL}
M.~Henneaux, V.~Lekeu, A.~Leonard, ``The action of the (free)
(4,0)-theory'', JHEP \textbf{01} (2018) 114, {\tt arXiv:01711.07448
[hep-th]}, [Erratum: JHEP \textbf{05} (2018) 105].

\bibitem{HLMP}
M.~Henneaux, V.~Lekeu, J.~Matulich, S.~Prohazka, ``The action the
(free) ${\cal N}=(3,1)$ theory in six spacetime dimensions'', JHEP
\textbf{06} (2018) 057, {\tt arXiv:1804.10125 [hep-th]}.

\bibitem{MSZ}
R.~Minasian, C.~Strickland-Constable, Y.~Zhang, ``On symmetries and
dynamics of exotic supermultiplets'', JHEP \textbf{01} (2021) 174,
{\tt arXiv:2007.08888 [hep-th]}.

\bibitem{BHHS}
Y.~Bertrand, S.~Hohenegger, O.~Holm, H.~Samtleben, ``Towards exotic
$6D$ supergravities'', Phys.~Rev. D \textbf{103} (2021) 046002, {\tt
arXiv:2007.11644 [hep-th]}.

\bibitem{Gu}
M.~Gunaydin, ``Unified non-metric (1,0) tensor-Einstein supergravity
theories and (4,0) supergravity in six dimensions'', JHEP
\textbf{06} (2021) 081, {\tt arXiv:2009.01274 [hep-th]}.


\bibitem{OP}
V.~I.~Ogievetsky, I.~V.~Polubarinov, ``The notoph and its possible interactions'', Yadernaya Fizika (Soviet Journal Nuclear Physics), \textbf{4} (1967) 156;
reprinted in {\it Supersymmetries and Quantum Symmetries}, J.~Wess and E.~A.~Ivanov (Eds.), Springer, 1999, pp. 391-396.

\bibitem{KR}
M.~Kalb, P.~Ramond, ``Classical direct interstring action'', Phys.~Rev. D \textbf{9} (1974) 2273.

\bibitem{CSc}
E.~Kremmer, J.~Scherk, ``Spontaneous dynamical breaking of gauge symmetry in dual models'', Nucl.~Phys., B \textbf{72} (1974) 117. 

\bibitem{GScO}
F.~Glozzi, J.~Scherk, D.~I.~Olive, ``Supersymmetry, supergravity theories and the dual spinor model'', Nucl.~Phys. B \textbf{122} (1977) 253.

\bibitem{Iv}
E.~A.~Ivanov, ``Gauge Fields, Nonlinear Realizations, Supersymmetry'', Phys.~Part.~Nucl., \textbf{47} (2016) 508, {\tt arXiv:1604.01379 [hep-th]}.

\bibitem{KUMM}
N.~Kummer, ``On the theory of particles of spin 1'', Helv.~Phys.~Acta, \textbf{33} (1960) 829. 

\bibitem{KR}
S.~M.~Kuzenko, E.~S.~N.~Raptakis, ``Covariant quantization of tensor
multiplet models'', JHEP \textbf{09} (2024) 182, {\tt
arXiv:2406.01176 [hep-th]}.

\bibitem{S1}
A.~S.~Schwarz, ``The partition function of degenerate quadratic
functionals and Ray-Singer invariants'', Lett.~Math.~Phys.
\textbf{2} (1978) 247.

\bibitem{S2}
A.~S.~Schwarz, ``The partition function of a degenerate
functional'', Commun.~Math.~Phys. \textbf{67} (1979) 1.

\bibitem{Sig}
W.~Siegel, ``Hidden ghosts'', Phys.~Lett. B \textbf{93} (1980) 170.

\bibitem{SN}
E.~Sezgin, P.~van Nieuwenhuizen, ``Renormalizability properties of
antisymmetric tensor field coupled to gravity'', Phys.~Rev. D
\textbf{22} (1980) 179.

\bibitem{HKO}
H.~Hata, T.~Kugo, N.~Ohta, ``Skew-symmetric tensor gauge field
theory dynamically realized in the QCD U(1) channel'', Nucl.~Phys. B
\textbf{178} (1981) 527.

\bibitem{Obukh}
Y.~N.~Obukhov, ``The geometrical approach to antisymmetric tensor
field theory'', Phys.~Lett, B \textbf{109} (1982) 195.

\bibitem{FT}
D.~Z.~Freedman, P.~K.~Townsend, ``Antisymmetric tensor gauge theories
and nonlinear sigma models'', Nucl.~Phys. B \textbf{177} (1981) 282.

\bibitem{TB}
J.~Thierry-Mieg, L.~Baulieu, ``Covariant quantization of nonabelian
antisymmetric tensor gauge theories'', Nucl.~Phys. B \textbf{228}
(1983) 259.

\bibitem{FrTs}
E.~S.~Fradkin, A.~A.~Tseytlin, ``Quantum equivalence of dual field
theories'', Ann.~Phys. \textbf{162} (1985) 31.

\bibitem{AGM1}
S.~P.~de Alwis, M.~T.~Grisaru, L.~Mazincescu, ``Unitarity in
antisymmetric tensor gauge theories'', Phys.~Lett. B \textbf{190}
(1987) 122. 

\bibitem{AF}
A.~A.~Slavnov, S.~A.~Frolov, ``Quantization of non-abelian
antisymmetric tensor field'', Theor.~Math.~Phys. \textbf{75} (1988)
470.

\bibitem{BK1}
I.~L.~Buchbinder, S.~M.~Kuzenko, ``Quantum equivalence of the
Freedman-Townsend model and the principal chiral $\sigma$-model'',
\text{unpublished} (1987), {\tt arXiv:2405.16782 [hep-th]}.

\bibitem{AGM2}
S.~P.~de Alwis, M.~T.~Grisaru, L.~Mazincescu, ``Quantization and
unitary in antisymmetric tensor gauge theories'', Nucl.~Phys. B
\textbf{303} (1988) 57.

\bibitem{BG}
C.~Battle, J.~Gomis, ``Lagrangian and Hamiltonian BRST structures of
the antisymmetric tensor gauge theory'', Phys.~Rev. D \textbf{38}
(1988) 1179. 

\bibitem{N}
N.~K.~Nielsen, ``Quantum equivalence of four-dimensional nonlinear
sigma-model and antisymmetric tensor model'', Nucl.~Phys. B
\textbf{332} (1990) 391.

\bibitem{GPS}
J.~Gomis, J.~Paris, S.~Samuel, ``Antibracket, antifields and
gauge theory quantization'', Phys. Repts. \textbf{259} (1995), 1,
{\tt arXiv:hep-th/9412228 [hep-th]}.

\bibitem{S1}
W.~Siegel, ``Gauge spinor superfield as a scalar multiplet'',
Phys.~Lett. B \textbf{85} (1979) 333.

\bibitem{G}
S.~J.~Gates, ``Super $p$-form gauge superfields'', Nucl.Phys. B
\textbf{184} (1981) 381.

\bibitem{GNSZ}
M.~T.~Grisaru, N.~K.~Nielsen, W.~Siegel, D.~Zanon, ``Energy-momentum
tensors, supercurrents, (super)traces and quantum equivalence'',
Nucl.~Phys. B \textbf{247} (1984) 157.

\bibitem{BK2}
I.~L.~Buchbinder, S.~M.~Kuzenko, ``Quantization of the classically
equivalent theories in the superspace of simple supergravity and
quantum equivalence'', Nucl.~Phys. B \textbf{308} (1988) 162.


\bibitem{BuKu}
I.~L.~Buchbinder, S.~M.~Kuzenko, ``Ideas and Methods of
Supersymmetry and Supergravity or a Walk Through Superspace'', IOP
Publishing, 1998.

\bibitem{FINN}
K.~Furita, T.~Inami, H.~Nakajima, M.~Nitta, ``Supersymmetric
extension of nonAbelian scalar tensor duality'', Prog.~Theor.~Phys.
\textbf{106} (2001) 851, {\tt arXiv:hep-th/0106138}.

\bibitem{FV}
E.~S.~Fradkin, G.~Vilkovisky, ``Quantization of relativistic
systems with constraints'', Phys.~Lett. B \textbf{55} (1975) 224.


\bibitem{BF}
I.~A.~Batalin, E.~S.~Fradkin, ``Relativistic S-matrix of dynamical
systems with boson and fermion constraints'', Phys.~Lett. B
\textbf{69} (1977) 309


\bibitem{Henn}
M.~Henneaux, C.~Teitelboim, ``Quantization of gauge systems'',
Princeton Univ. Press, 1992.


\bibitem{Bat1}
I.~A.~Batalin, G.~A.~Vilkovisky, ``Gauge Algebra and Quantization'',
Phys.~Lett. B \textbf{102} (1981) 27.

\bibitem{Bat2}
I.~Batalin, G.~Vilkovisky, ``Quantization of Gauge Theories with
Linearly Dependent Generators'', Phys.~Rev. D \textbf{28} (1983)
2567, [Erratum: Phys.~Rev. D \textbf{30} (1984) 508].

\bibitem{BBKN1}
A.~O.~Barvinsky, I.~L.~Buchbinder, V.~A.~Krykhtin, D.~V.~Nesterov, ``Adjustment
of Faddeev-Popov quantization to reducible gauge theories:
Antisymmetric tensor fermion in $AdS_d$ space'', Phys.~Rev. D
\textbf{112} (2025) 065021, {\tt arXiv:2509.01863 [hep-th]}.

\bibitem{BBKN2}
A.~O.~Barvinsky, I.~L.~Buchbinder, V.~A.~Krykhtin, D.~V.~Nesterov,
``Covariant quantization of totally antisymmetric tensor-spinor
field in $AdS_d$'', Phys.~Rev. D \textbf{113} (2026) 045022, {\tt
arXiv:2509.01863 [hep-th]}.

\bibitem{LZ}
V.~Lekeu, Y.~Zhang, ``On the quantization and anomalies of
antisymmetric tensor-spinors'', JHEP \textbf{11} (2021) 078, {\tt
arXiv:2109.03963 [hep-th]}.

\bibitem{FP}
L.~D.~Faddeev, V.~N.~Popov, ``Feynman Diagrams for the Yang-Mills Field'', Phys. Lett. B \textbf{25} (1967) 29. 


\bibitem{Buchbinder:2008kw}
I.~L.~Buchbinder, V.~A.~Krykhtin, L.~L.~Ryskina,
``BRST approach to Lagrangian formulation of bosonic totally antisymmeric tensor fields in curved space'',
Mod. Phys. Lett. A \textbf{24} (2009), 401-414
{\tt arXiv:0810.3467 [hep-th]}.

\end{thebibliography}
\end{document}